\documentclass[lettersize,journal]{IEEEtran}
\usepackage{amsmath,amsfonts}
\usepackage{algorithmic}
\usepackage{algorithm}
\usepackage{array}
\usepackage[caption=false,font=normalsize,labelfont=sf,textfont=sf]{subfig}
\usepackage{textcomp}
\usepackage{stfloats}
\usepackage{url}
\usepackage{verbatim}
\usepackage{graphicx}
\usepackage{hyperref}
\usepackage{cite}
\usepackage{array, makecell}
\usepackage{mathtools}

\usepackage[font=small,labelfont=bf]{caption}
\newcommand\halfwidth{1.0} 

\begin{document}

\title{Defending Adversarial Attacks on Deep Learning-Based Power Allocation in Massive MIMO Using Denoising Autoencoders}

\author{Rajeev Sahay*,~\IEEEmembership{Graduate Student Member,~IEEE,}
        Minjun Zhang*,\\~David~J.~Love,~\IEEEmembership{Fellow,~IEEE}, and Christopher~G.~Brinton,~\IEEEmembership{Senior~Member},~IEEE

\thanks{Code is publicly available at \url{https://github.com/Jess-jpg-txt/DAE_for_adv_attacks_in_MIMO}.}
\thanks{*R. Sahay and M. Zhang contributed equally to this work.}  
\thanks{R. Sahay is with the Autonomy and Undersea Systems Division, Saab, Inc., West Lafayette, IN, 47907. E-mail: Rajeev.Sahay@saabinc.com.}%
\thanks{R. Sahay and M. Zhang were with the Elmore Family School of Electrical and Computer Engineering, Purdue University, West Lafayette, IN, 47907.}%
\thanks{D. J. Love and C. G. Brinton are with the Elmore Family School of Electrical and Computer Engineering, Purdue University, West Lafayette, IN, 47907 USA. E-mail: \{djlove,cgb\}@purdue.edu.}
\thanks{This work was supported in part by the Office of Naval Research (ONR) under grants N00014-21-1-2472 and N00014-22-1-2305, in part by the National Science Foundation (NSF) under grants CNS-2146171, EEC-1941529, CNS-2225577, and CNS-2212565, and in part by the Defense Advanced Research Projects Agency (DARPA) under grant D22AP00168-00.}}



\maketitle

\begin{abstract}

Recent work has advocated for the use of deep learning to perform power allocation in the downlink of massive MIMO (maMIMO) networks. Yet, such deep learning models are vulnerable to adversarial attacks. In the context of maMIMO power allocation, adversarial attacks refer to the injection of subtle perturbations into the deep learning model's input, during inference (i.e., the adversarial perturbation is injected into inputs during deployment after the model has been trained) that are specifically crafted to force the trained regression model to output an infeasible power allocation solution. In this work, we develop an autoencoder-based mitigation technique, which allows deep learning-based power allocation models to operate in the presence of adversaries without requiring retraining. Specifically, we develop a denoising autoencoder (DAE), which learns a mapping between potentially perturbed data and its corresponding unperturbed input. We test our defense across multiple attacks and in multiple threat models and demonstrate its ability to (i) mitigate the effects of adversarial attacks on power allocation networks using two common precoding schemes, (ii) outperform previously proposed benchmarks for mitigating regression-based adversarial attacks on maMIMO networks, (iii) retain accurate performance in the absence of an attack, and (iv) operate with low computational overhead.

\end{abstract}


\section{Introduction}







\IEEEPARstart{M}{assive} multiple-input-multiple-output (maMIMO) systems have incurred exciting developments in recent years, both in theory \cite{mimo_survey} and in practice \cite{mamimo1}. A maMIMO system uses a wireless network topology where each base station (BS) is equipped with a large number of antennas to serve a multitude of user equipments (UEs) by using the spatial degrees-of-freedom \cite{mimo}. One major task of the BS in maMIMO networks is feasible power allocation to each serviced UE. Accurate power allocation from the BS to each UE is vital for efficient communication (e.g., to increase the sum rate  and reduce operational costs) in maMIMO networks. Although the equal power allocation policy can equally distribute power to all UEs, it is far from optimal and cannot always meet the power allocation needs of each UE simultaneously. 
Thus, power allocation algorithms that can meet each UE's power needs while staying within the BS's total power budget are required. 

A variety of maMIMO power allocation algorithms have been developed, given a specific precoder, using convex optimization and geometric programming \cite{trad_pa_1,trad_pa_2,trad_pa_3,trad_pa_4,trad_pa_5,trad_pa_6}. Yet, such methods often computationally costly, which cannot always be afforded in real-time systems \cite{beyond_5g}. To address this, recent work has demonstrated the effectiveness of power allocation in maMIMO networks, given a specific precoder, using deep learning, which can attain near-optimal performance by using the positions of UEs as input into a deep neural network (DNN) \cite{dl_in_mimo}. Moreover, deep learning is capable of improving the complexity-performance trade-off in comparison to traditional optimization-based methods while also eliminating the need to compute any statistical averages usually needed in standard methods \cite{dl_in_mimo2,dl_in_mimo3,dl_in_mimo4}. 

The utilization of deep learning for wireless systems, however, has introduced a new class of malicious attacks. Specifically, deep learning-based power allocation models in maMIMO have been shown to be susceptible to gradient-based \emph{adversarial evasion attacks} -- subtle perturbations injected into a DNN's input at inference to induce erroneous behavior at the output \cite{adv_mimo1,adv_mimo2,adv_mimo3,adv_mimo4}. In the context of maMIMO power allocation, erroneous model behavior refers to the adversarial perturbation inducing an infeasible power allocation to the UEs. Such attacks cause the DNN to unfairly allocate resources to certain UEs, thus hindering the effectiveness and efficiency of deep learning in maMIMO networks. Contrary to deep learning-based power allocation, traditional optimization-based power allocation techniques are not hindered specifically by adversarial attacks (although other perturbations may degrade their performance \cite{mimo_spoof,mimo_atk2,mimo_atk3}) since there is no model that can be used to craft a gradient-based attack. Despite this advantage, optimization-based methods still incur high computational overhead, which may not always be affordable in real-time \cite{dl_in_mimo,dl_in_mimo2,dl_in_mimo3}. Thus, mitigating adversarial attacks in DNN-based power allocation is needed to allow for both the computational efficiency and model robustness of deep learning in maMIMO systems. 


In this work, we consider a denoising autoencoder (DAE) framework for mitigating the effects of adversarial attacks instantiated on DNN regression models for power allocation in maMIMO networks. We begin by developing our DAE, which is a layer-based DNN architecture trained to map perturbed adversarial samples to their corresponding clean (non-adversarial) representations. In addition, the DAE is also trained to map clean, unperturbed data to itself in order to retain performance on non-adversarial samples. During inference, the received sample is first forward propagated through the DAE, and the DAE output is then inputted into the DNN regression model as the BS. We evaluate, in terms of both robustness and computational efficiency, the feasibility of our method against multiple attack generation algorithms on two precoding schemes. 

Formally, our contributions of this work can be summarized as follows: 
\begin{enumerate}

    
    \item \textbf{DAE output mitigation strategy against adversarial attacks} (Sec. III, V, Sec. VI-B, and Sec. VI-C): We develop a DAE architecture capable of filtering adversarial perturbations out of maMIMO input data on deployed DNNs. At test time, we show that forward propagating the input (i.e., UE positions) through the DAE prior to inputting it into the DNN significantly mitigates the effects of adversarial attacks. 
    
    
    
    
    \item \textbf{Effectiveness of defense in a multitude of environments } (Sec. IV, VI-A -- Sec. VI-C): We demonstrate the effectiveness of our defense in both semi-white box and black box threat models on maMIMO networks using both maximum ratio (MR) and multicell-minimum mean square error (MMMSE) precoding schemes. 
    
    \item \textbf{Robustness against first order $l_{\infty}$-bounded adversarial attacks} (Sec. VI-B and Sec. VI-C): We show that our proposed DAE defense provides robustness against multiple first-order $l_{\infty}$-bounded adversarial attacks on regression-based power allocation DNNs in multiple threat models and on multiple precoding schemes. 
    
    \item \textbf{Computational Feasibility} (Sec. VI-D): We quantify the computational speedup provided by our proposed method over comparative benchmarks. Specifically, we find that our proposed framework does not incur higher computational overhead as the parameter space of the employed DNN increases.  
    
\end{enumerate}

The remainder of the paper is organized as follows. Sec. \ref{related_work} discusses prior work on adversarial attacks and defenses instantiated on deep learning-based massive MIMO power allocation as well as other relevant deep learning-based communications tasks. Sec. \ref{dl_for_pa} describes our considered setup for performing power allocation in massive MIMO networks using deep learning. Sec. \ref{adv_atk_for_pa} discusses our adversarial attack formulation, considered adversarial attacks, and the adversary's various knowledge levels. Sec. \ref{dae_sec} details our denoising autoencoder (DAE) defense and implementation details. Finally, Sec. \ref{results_sec} presents the results of our methodology's empirical evaluation. We present concluding remarks in Sec. \ref{conclusion_sec}. 

\section{Related Work} \label{related_work}


Deep learning has recently achieved cutting edge performance in the domain of wireless communications. The majority of these applications have consisted of classification problems including, but not limited to, automatic modulation classification \cite{amc_dl}, sensing and localization \cite{sl1,sl2}, and channel estimation \cite{chan_est}. Such tasks have extensively been shown to be vulnerable to adversarial attacks \cite{adv_in_wc,iot_adv_atk}, with a particular emphasis on automatic modulation classification \cite{amc_adv_atk1,amc_adv_atk2,amc_adv_atk3,amc_adv_atk4,amc_adv_atk5}. Similarly, previous studies have proposed methods to craft effective wireless eavesdroppers \cite{adversary_atk}, while other works have examined circumventing such eavesdroppers by intentionally transmitting adversarially perturbed data that the receiver can subtract from the observed signal \cite{eve1,eve2,eve3}. Our work can be considered orthogonal to such approaches in that our proposed framework is (i) based on regression and not classification and (ii) aiming to directly induce erroneous behavior at the BS and not fool an eavesdropper. 

As a result of the susceptibility of wireless signal classifiers to adversarial perturbations, several defenses have been proposed such as autoencoder pre-training \cite{ae_pt}, detection and rejection \cite{adv_amc_detection}, and assorted deep ensembles \cite{ade}. Despite the strides made in defending signal classifiers from adversarial attacks, these methods are difficult to directly adopt for defending adversarial attacks on regression DNNs for the task of power allocation in the downlink of maMIMO because they are specifically tailored for classification tasks. In this work, we address this shortcoming by developing a defense that is capable of mitigating adversarial attacks on DNNs used for regression tasks in maMIMO networks.


Furthermore, in addition to the host of defenses studied in classification-based wireless communications, defenses against adversarial attacks have been investigated to a large extent in computer vision \cite{cv_review}. In particular, DAEs have shown success in mitigating adversarial attacks for computer vision tasks such as image classification \cite{dae,dunet}. However, DAEs have not been explored for mitigating adversarial attacks in regression-based computer vision tasks. For mitigating adversarial attacks in regression-based maMIMO networks, adversarial training \cite{pgd_adv_trn,fgsm,adv_trn_sahay} has been shown to effectively reduce the adversary's success rate in white-box \cite{mimo_adv_atk_def} and black box settings \cite{mimo_adv_trn}. Despite its success, adversarial training remains vulnerable to high-bounded perturbations and, further, suffers from overfitting to adversarial examples, which, in turn, reduces performance on unperturbed data \cite{adv_trn_ovft}. Our proposed DAE, on the other hand, remains effective at high perturbation magnitudes without significantly reducing baseline prediction performance. 

Lastly, it is important to note that prior work has demonstrated the ineffectiveness of pre-processing defense mechanisms in defending adversarial attacks in computer vision applications \cite{obf_grd}. However, contrary to computer vision, adversaries in the context of maMIMO networks will rarely have complete knowledge of each processing module employed at the BS \cite{adv_mimo3,bb_justification}. Thus, pre-processing defense mechanisms, as we will see, serve as effective mitigation techniques in the context of wireless communications, where the adversary is typically knowledge-limited while operating. 






\section{Deep Learning for Power Allocation} \label{dl_for_pa}

In this section, we describe the maMIMO power allocation objective formulated for the downlink and describe how to perform power allocation using DNNs.



\subsection{Downlink Power Allocation}
We begin by considering a maMIMO system with $L$ cells, where each cell contains a BS with $M$ antennas and $K$ UEs per cell. The downlink signal transmitted from BS $j$ is given by 
\begin{equation} \label{dl_sig}
    \sum_{k=1}^{K} \mathbf{f}_{jk} \varsigma_{jk},
\end{equation}
where $\varsigma_{jk} \sim \mathcal{CN}(0, \rho_{jk})$ is the downlink data signal for UE $k$ in cell $j$, and $\mathbf{f}_{jk} \in \mathbb{C}^{M}$ is the precoding vector, which determines the directivity of the transmission. Furthermore, $||\mathbf{f}_{jk}||^{2} = 1$ and thus $\rho_{jk}$ corresponds to the transmit power. 
Following (\ref{dl_sig}) as the transmitted downlink signal, we denote the received signal of UE $k$ in cell $j$ as
\begin{align} \label{rec_sig} 
    \mathbf{y}_{jk} = \underbrace{(\mathbf{h}_{jk}^{j})^{\text{H}} \mathbf{f}_{jk} \varsigma_{jk}}_\text{Desired signal} + & \underbrace{\sum_{i \neq k}^{K}  (\mathbf{h}_{jk}^{j})^{\text{H}} \mathbf{f}_{ji} \varsigma_{ji}}_\text{Intra-cell Interference} + \nonumber \\
    & \underbrace{\sum_{l \neq j}^{L} \sum_{i = 1}^{K}  (\mathbf{h}_{jk}^{l})^{\text{H}} \mathbf{f}_{li} \varsigma_{li}}_\text{Inter-cell interference} + \underbrace{n_{jk}}_\text{AWGN},
\end{align}
where $(\cdot)^{\text{H}}$ denotes the Hermitian operator, $\mathbf{h}_{jk}^{l}$ is the channel between BS $l$ and UE $k$ in cell $j$ (i.e., the superscript of $\mathbf{h}_{jk}^{l}$ corresponds to the transmit BS and the subscripts identify the cell and index of the receive UE), and $n_{jk} \sim \mathcal{CN}(0, \sigma^{2})$ is complex additive white Gaussian noise (AWGN) with variance $\sigma^{2}$. 

We assume that the uplink and downlink channels are reciprocal within a coherence block and the UE can thus approximate the precoded channel through its expected value, $\mathbb{E}[(\mathbf{h}_{jk}^{j})^{\text{H}} \mathbf{f}_{jk}]$, where $\mathbb{E}[\cdot]$ denotes the expectation operator. This is a deterministic value that the UE has access to (as assumed in \cite{mamimo_book}). The remaining precoded channels in (\ref{rec_sig}) (i.e., the inter-cell and intra-cell interference) are treated as realizations of random variables whose expectations are taken at the UE with respect to different channel realizations. Each of these quantities are then used to calculate the signal-to-interference-plus-noise ratio (SINR) of (\ref{rec_sig}) in the downlink for UE $k$ in cell $j$. The SINR, which is a deterministic scalar, is given by (see \cite{mamimo_book})
\begin{equation} \label{sinr}
    \gamma_{jk}^\text{dl} = \frac{\rho_{jk} a_{jk}} {\sum_{l=1}^{L} \sum_{i=1}^{K} \rho_{li} b_{lijk} + \sigma^{2}}, 
\end{equation}
where 
\begin{equation} \label{ch_gain}
    a_{jk} = |\mathbb{E}[\mathbf{f}_{jk}^{\text{H}} \mathbf{h}_{jk}^{j}]|^{2}, 
\end{equation}
and
\begin{equation} \label{int_gain}
    b_{lijk} = 
    \begin{cases}
      \mathbb{E}[|\mathbf{f}_{jk}^{\text{H}} \mathbf{h}_{jk}^{l}|^{2}], & (l, i) \neq (j, k)\\
      \mathbb{E}[|\mathbf{f}_{li}^{\text{H}} \mathbf{h}_{jk}^{l}|^{2}] - |\mathbb{E}[\mathbf{f}_{li}^{\text{H}} \mathbf{h}_{jk}^{j}]|^{2}, & (l, i) = (j, k) 
    \end{cases}, 
\end{equation}
are the expected channel and interference gains, respectively.  

Selecting the optimal precoder, $\mathbf{f}_{jk}$, is a challenging task since (\ref{sinr}) depends on the precoders of all UEs in the maMIMO network. Thus, following prior work (see Sec. II-C in \cite{dl_in_mimo}), we use the UL-DL duality and design the precoder based on the UL combining vector (used to detect the UL signal transmitted by the UE), where we consider MR combining \cite{mr_precoding} and MMMSE combining \cite{mamimo2}. Thus, throughout this work, we will consider the performance of our framework when the precoder is based on both MR and MMMSE combining. 



Using the above definitions of SINR, the optimal power allocation to UEs is the solution of the max-product policy given by \cite{mamimo_book}
\begin{subequations} \label{p_all_opt}
\begin{align}
    \underset {\rho_{jk: \forall j, k}} {\text{max}} \quad & \prod_{j=1}^{L} \prod_{k=1}^{K} \gamma_{jk}^{\text{dl}} \\
    \text{s. t.} \quad &  \hspace{0.5mm} \sum_{k=1}^{K} \rho_{jk} \leq P_{\text{max}}^{\text{dl}}, j \in [1, L], 
\end{align}
\end{subequations}
where $P_{\text{max}}^{\text{dl}}$ is the downlink transmit power budget. The optimal power allocation solutions to (\ref{p_all_opt}) are obtained by computing (i) the large-scale fading coefficients and channel correlation matrices, (ii) the random channel vectors, (iii) $\mathbf{f}_{jk}$, $\{a_{jk}\}$, and $\{b_{jk}\}$ and, lastly, (iv) allocating power to UEs to satisfy (\ref{p_all_opt}) \cite{dl_in_mimo,mamimo_book}. This algorithm is implemented using geometric programming, which requires polynomial complexity. However, such complexity may not be affordable in real-time systems.By calculating the optimal power allocation for UEs at various locations ahead of time, a mapping between UE positions and their optimal power allocation can be learned for faster real-time power allocation prediction. 

\subsection{Deep Learning Implementation}
\label{sec:dl-imp}

\begin{table} [t]
\small
\caption{The DNN architecture of Model 1 \label{model1}}
\centering
\begin{tabular}{c | c | c } 
\centering 
\makecell{Layer} & \makecell{Units } & \makecell{Activation \\ Function } \\ 
\hline
Input & $KL$ & - \\
Layer 1 & 256 & ELU \\
Layer 2 & 128 & ELU \\
Layer 3 & 64 & ELU \\
Layer 4 & 64 & ELU \\
Layer 5 & $K$ & ELU \\
Output & $K+1$ & Linear \\

\hline

\end{tabular}
\end{table}

\begin{table} [t]
\small
\caption{The DNN architecture of Model 2 \label{model2}}
\centering
\begin{tabular}{c | c | c } 
\centering 
\makecell{Layer} & \makecell{Units } & \makecell{Activation \\ Function } \\ 
\hline
Input & $KL$ & - \\
Layer 1 & 512 & ELU \\
Layer 2 & 256 & ELU \\
Layer 3 & 128 & ELU \\
Layer 4 & 128 & ELU \\
Layer 5 & $K$ & ELU \\
Output & $K+1$ & Linear \\

\hline

\end{tabular}
\end{table}

The objective of employing a DNN at the BS is to (i) learn the optimal power allocation, given by the max-product policy, in the downlink of a maMIMO network and (ii) reduce the latency incurred from solving (\ref{p_all_opt}) when UE positions change in real-time. To do this, training data is first generated by solving the formulation given in (\ref{p_all_opt}) using geometric programming for UEs positioned at various locations throughout the considered cells (assuming intra-cell UE movement, as considered in \cite{dl_in_mimo,mimo_adv_trn,adv_mimo1,adv_mimo3,adv_mimo4}, but not inter-cell UE movement). Then, a DNN is trained to learn the mapping between UE positions and the optimal power allocations obtained from solving (\ref{p_all_opt}). Both the calculation of (\ref{p_all_opt}) and the DNN training are done prior to implementation, and, thus, their incurred overhead does not contribute to real-time computational costs. During implementation, the DNN significantly improves latency because it (i) eliminates the need to calculate $\{a_{jk}\}$ and $\{b_{lijk}\}$ and (ii) also eliminates the need to solve (\ref{p_all_opt}) when the $K$ UEs have moved within a cell. As a result, the overall runtime during deployment can be reduced by up to $99.9\%$ \cite{dl_in_mimo3}. 

Although generating the training data requires polynomial computational complexity, since (\ref{p_all_opt}) has to be solved for each training sample, this can be done offline, and thus higher complexity can be afforded since real-time constraints do not apply. Specifically, UE positional data and their corresponding optimal power allocations obtained from solving (\ref{p_all_opt}) are assumed to be collected and stored in a central server where the DNNs are trained (corresponding to the offline development) before the trained models are deployed at the BSs for real-time power allocation (corresponding to the online deployment). Thus, the trained DNN eliminates the need to re-calculate the needed power for different sets of UEs as they move around the cells, since their new power requirement can be predicted by forward propagating the UE locations through the trained DNN at the BS, leading to significantly improving latency.

Formally, we define the $n^{\text{th}}$ input to the DNN as $\mathbf{x}(n) \in \mathbb{R}^{2KL}$, where $\mathbf{x}(n) = [x_{1l}, y_{1l}, \cdots, x_{KL}, y_{KL}]$ represents the geographical coordinates, in a Euclidean space, of each of the $(KL)$ UEs. The corresponding DNN target output is denoted by $\pmb{\rho}_{j}(n) \in \mathbb{R}^{K + 1}$, which is the optimal power allocation for each UE in cell $j$ obtained for training by solving (\ref{p_all_opt}) with knowledge of (\ref{ch_gain}) and (\ref{int_gain}). Note that, in addition to the power allocation for each UE, $\pmb{\rho}_{j}(n)$ also contains the power constraint $\sum_{k=1}^{K} \rho_{jk}$, where $\sum_{k=1}^{K} \rho_{jk} = P_{\text{max}}^{\text{dl}}$ and $P_{\text{max}}^{\text{dl}}$ is always fixed, such that $\pmb{\rho}_{j}(n) = [\rho_{j1}, \ldots, \rho_{jK}, \sum_{k=1}^{K} \rho_{jk}]$ \cite{adv_mimo1}, in order to improve the DNN estimation ability during training and inference. In addition, each BS learns the mapping between UE positions, $\mathbf{x}(n)$, and power allocation, $\pmb{\rho}_{j}(n)$, for the individual cell $j$. 

Next, we define $\mathcal{X}_{\text{tr}} = \{\mathbf{x}(n), \pmb{\rho}_{j}(n); n=1, \ldots,N\}$ as the training dataset containing $N$ samples. Using this dataset, at each BS, we train a DNN, $f^{l}(\cdot, \pmb{\theta}_{l}): \mathbb{R}^{2KL} \rightarrow \mathbb{R}^{K+1}$, parametertized by $\pmb{\theta}_{l}$, to map each input sample, $\mathbf{x}(n)$, (i.e., the position of each UE among all $L$ cells) to the appropriate power allocation vector, $\pmb{\rho}_{j}$, for each UE in the cell. Here, the inputs only contain the positional information of the UEs; the target label, containing the optimal power allocation vector obtained from (\ref{p_all_opt}), capture the remaining effects of the channel. The DNN takes input $(\cdot)$, and outputs the predicted power, $\hat{\pmb{\rho}}_{j}$, to allocate to each UE in its corresponding cell, i.e., the estimated power allocation by the DNN to the UEs serviced by BS $j$ is given by $\hat{\pmb{\rho}}_{j}(n) = f^{l}(\mathbf{x}(n), \pmb{\theta}_{l})$. During evaluation, we employ the dataset $\mathcal{X}_{\text{te}} = \{\mathbf{x}(b), \pmb{\rho}_{j}(b); b=1, \ldots,B\}$, where $\mathcal{X}_{\text{tr}} \cap \mathcal{X}_{\text{te}} = \emptyset$ (i.e., no samples overlap in the training and evaluation datasets).


We evaluate our methodology on two considered DNN architectures. The first model, referred to as Model 1, is a custom architecture, which we found capable of learning the desired mapping between UEs and optimal power allocation. The second model, Model 2, is adopted from prior work \cite{dl_in_mimo,adv_mimo1} that has shown it to be effective for power allocation in maMIMO systems. Both considered models are shown in Table \ref{model1} and Table \ref{model2}, respectively. Each model is a fully connected neural network (FCNN) comprised of individual units, where each unit contains a set of trainable weights, whose input dimensionality is equal to the number of units in the preceding layer, and the number of units in each layer is an adjustable hyper-parameter. The output of the $i^{\text{th}}$ layer of the DNN, prior to applying the activation, is given by $\mathbf{z}^{(i)} = \mathbf{W}^{(i)}\mathbf{o}^{(i-1)} + \mathbf{b}^{(i)}$, 
\noindent where $\mathbf{W}^{(i)}$ is the weight matrix of layer $i$, $\mathbf{o}^{(i-1)}$ is the input vector to layer $i$ containing the output from the previous layer, and $\mathbf{b}^{(i)}$ is a vector of threshold biases. 


Each intermediate layer in both Model 1 and Model 2 applies the unit-wise exponential linear unit (ELU) activation function, thus resulting in the output of each unit in hidden layer $i$ given by $\mathbf{o}^{(i)}(p) = \nu(\mathbf{z}^{(i)}(p))$ where 
\begin{equation} \label{elu}
    {\nu}\big{(} \mathbf{z}^{(i)}(p) \big{)}  = 
    \begin{cases}
      \mathbf{z}^{(i)}(p), & \mathbf{z}^{(i)}(p) \geq 0\\
      \text{exp}\big{(}\mathbf{z}^{(i)}(p)\big{)} - 1, & \mathbf{z}^{(i)}(p) < 0
    \end{cases},  
\end{equation}
and the output of the final layer applies the linear activation function resulting in ${\nu}\big{(}  \mathbf{z}^{(i)}(p) \big{)}  =  \mathbf{z}^{(i)}(p)$,
where $\nu(\cdot)$ denotes the activation function, $\mathbf{z}^{(i)}(p)$ and $\mathbf{o}^{(i)}(p)$ denote the $p^{\text{th}}$ unit of layer $i$ prior to and after applying the activation, respectively, and $\mathbf{z}^{(i)}$ and $\mathbf{o}^{(i)}$ are the output vectors of layer $i$ prior to and after applying the activation, respectively. 

Using $\mathcal{X}_{\text{tr}} = \{\mathbf{x}(n), \pmb{\rho}_{j}(n); n=1, \ldots,N\}$, both Model 1 and Model 2 are trained by minimizing the relative mean squared error, given by 
\begin{equation}
    \ell(\mathbf{x}(n), \pmb{\rho}_{j}(n), \pmb{\theta}_{l}) = \frac{1}{N} \sum_{n=1}^{N} \bigg{(}\frac {\pmb{\rho}_{j}(n) - \hat{\pmb{\rho}}_{j}(n)} {\pmb{\rho}_{j}(n)}\bigg{)}^{2},
\end{equation}
using stochastic gradient descent (SGD) so that the model parameters on epoch $t +1$ are updated according to\footnote{In practice, mini-batches are used to accelerate training.}
\begin{equation} \label{sgd}
    \pmb{\theta}_{l}^{t+1} = \pmb{\theta}_{l}^{t} - \eta \frac{\partial \ell(\mathbf{x}(n), \pmb{\rho}_{j}(n), \pmb{\theta}_{l})}{\partial \pmb{\theta}_{l}},
\end{equation}
where $\eta$ is the model learning rate and the parameters for the first epoch, i.e., $\pmb{\theta}_{l}^{0}$, are randomly initialized. Rather than using a fixed learning rate, our training process starts with a learning rate of $10^{-1}$ and gradually reduces to $10^{-4}$ as the model gets closer to convergence. When multiple consecutive epochs yield indistinguishable relative mean squared errors for their validation sets, we reduce the learning rate by a factor of 0.1 to ensure the model takes smaller steps towards convergence. We consider this training process for both Model 1 and Model 2 in the cases of both MR and MMMSE precoding.

\section{Adversarial Attacks on Power Allocation} \label{adv_atk_for_pa}

In this section, we will begin by introducing the adversarial attack formulation on the power allocation DNNs described in Sec. III. Then, we will discuss two approaches of crafting adversarial attacks for inducing infeasible power allocation solutions by the DNN. Finally, we will discuss the adversarial threat model, where we will state the various knowledge levels we consider the adversary to operate in and specify the available system information known by the adversary. 

\subsection{Attack Formulation}

Adversarial attacks on DNNs can be crafted in a multitude of ways (as further discussed in Sec. \ref{sec:fgsm} and \ref{sec:pgd}) by introducing subtle perturbations into an input sample. In this work, we assume that the BS first obtains each UE's positional information for power allocation. After obtaining the UE positions at the BS, and prior to its input to the DNN, an adversary crafts and injects a perturbation into the DNN input sample. Physically, an attack crafted to perturb an input in this fashion corresponds to changing the UEs current position to a new position by a small amount (on the scale of tens of meters). Such an attack can be instantiated via spoofing attacks by using global navigation satellite system (GNSS) receivers, where an adversary calculates a minimum distance perturbation that will result in erroneous outputs made by the DNN \cite{spoofing}. Furthermore, such attacks may also induce from position equivocation algorithms leading to inexact UE positional information at the BS.

Formally, the adversary aims to inject the $n^{\text{th}}$ sample with a perturbation, $\pmb{\delta}(n) \in \mathbb{R}^{2KL}$, thus yielding 
\begin{equation}
    \tilde{\mathbf{x}}(n) = {\mathbf{x}}(n) + \pmb{\delta}(n),
\end{equation}
where $\tilde{\mathbf{x}}(n)$ represents the perturbed adversarial sample. As the adversary is aiming to operate undetected, while simultaneously trying to degrade the DNN performance, an effective perturbation can be generated by solving
\begin{subequations} \label{adv_opt:all-lines}
\begin{align}
    \underset{\pmb{\delta}(n)}{\text{min}} \quad &  ||\pmb{\delta}(n)||_{\infty} \label{adv_opt:line_1} \\
    \text{s. t.} \quad &  \hspace{0.5mm}  \sum_{k=1}^{K} \hat{\rho}_{jk} > P_{\text{max}}^{\text{dl}} \label{adv_opt:line_2} \\ 
    \quad & ||\pmb{\delta}(n)||_{\infty} \leq \epsilon, \label{adv_opt:line_3} \\
     \quad & {\mathbf{x}(n)} + \pmb{\delta}(n) \in \mathbb{R}^{2KL} \label{adv_opt:line_4}, 
\end{align}
\end{subequations} 
where $||\cdot||_{\infty}$ is the $l_{\infty}$ norm and $\epsilon$ is the maximum allowed perturbation for each feature in $\mathbf{x}(n)$ (i.e., $||\tilde{\mathbf{x}}(n) - \mathbf{x}(n)||_{\infty} \leq \epsilon$), thus directly corresponding to the the positional displacement of the UE. Although (\ref{adv_opt:line_1}) specifically constrains the $l_{\infty}$ norm of $\pmb{\delta}(n)$, note that the adversary aims to inject its $\epsilon$-constrained perturbation into each feature (positional coordinate) of $\mathbf{x}(n)$ (i.e., $\pmb{\delta}(n) \in \mathbb{R}^{2KL}$). The ultimate objective of this design, from the adversary's perspective, is to force the DNN to output an infeasible power allocation solution as in (\ref{adv_opt:line_2}), while remaining in (i) an $\epsilon$-neighborhood of the original sample as in (\ref{adv_opt:line_3}) and (ii) the same dimensional space as the input as given by (\ref{adv_opt:line_4}). Furthermore, note that the perturbation is added after normalizing the sample and prior to inputting it to the DNN and, therefore, the normalization does not aid in mitigating the attack since the perturbation is added after normalization. 

Due to the excessive nonlinearity of the considered DNN architectures as well as the non-convex nature of the adversary's objective function, solving (\ref{adv_opt:all-lines}) directly is very challenging. Thus, several gradient-based adversarial attacks have been proposed in order to approximate a solution that ideally solves (\ref{adv_opt:all-lines}) in effective and computationally efficient ways \cite{grad1,grad2}. Since solutions to (\ref{adv_opt:all-lines}) are typically approximated, they may not always be effective (i.e., there may be designs of $\pmb{\delta}(n)$ for which certain constraints in (\ref{adv_opt:all-lines}) are not satisfied). Specifically, an adversarial attack is effective when the DNN outputs an infeasible power allocation to the UEs as a result of the perturbation and ineffective otherwise. Moreover, the adversary's primary objective is to force an infeasible output by the DNN and, thus, an effective approximation to (\ref{adv_opt:all-lines}) may not necessarily be realized when $||\pmb{\delta}(n)||_{\infty}$ is minimized.

As shown in (\ref{adv_opt:all-lines}), we consider the class of adversarial attacks bounded by the $l_{\infty}$ norm constraint. However, adversarial attacks can be constrained by other $l_{p}$ norms as well such as when $p = 0$, $p=1$, or $p=2$. In this work we focus exclusively on $l_{\infty}$ norm since the input to the DNN is the position of the UE in a Euclidean space, and thus the $l_{\infty}$ norm corresponds to the maximum positional displacement of the UE. Note that the distance of the UE displacement for an attack bound of $\epsilon$ is given by $d_{\epsilon} = \sqrt{2}\epsilon$ since the perturbation displaces the UE along both dimensions of the Euclidean space.

We consider two adversarial attack algorithms that approximate solutions to (\ref{adv_opt:all-lines}): the fast gradient sign method (FGSM) \cite{fgsm} and projected gradient descent (PGD) \cite{pgd}. The FGSM is a single step attack making it computationally efficient whereas PGD is iterative thus requiring higher complexity to implement but resulting in a more potent attack compared to the FGSM. Both attacks rely on adding a perturbation in the direction of some loss function's gradient to achieve their objective. In order to maximize the probability of crafting an effective attack, and following in line with prior works \cite{adv_mimo3,mimo_adv_trn}, we define the loss function used by the adversary to be
\begin{equation} \label{adv_loss}
    \mathcal{L}(\mathbf{x}(n), \pmb{\theta}) = \sum_{k=1}^{K} \hat{\rho}_{jk},
\end{equation}
where $\hat{\rho}_{jk}$ corresponds to the predicted power to be allocated to UE $k$ in cell $j$ on the input sample $\mathbf{x}(n)$. The adversary may either use the DNN deployed at the BS to measure this loss, or they may alternatively use a surrogate DNN. The DNN used by the adversary will depend on the amount of available system knowledge, which is further discussed in Sec. IV-D. Next, we will discuss how the adversary uses (\ref{adv_loss}) to craft $\pmb{\delta}(n)$ using the FGSM and PGD algorithms. Each considered attack is applied cell-wise, where the adversary aims to degrade performance of the DNN in the cell serviced by the respective BS. 





\subsection{Fast Gradient Sign Method (FGSM)}
\label{sec:fgsm}

The FGSM is a single step method that is given by 
\begin{equation} \label{delta_fgsm}
    \pmb{\delta}(n) = \epsilon \hspace{0.5mm} \text{sign}\bigg{(}\nabla_{\mathbf{x}(n)}\mathcal{L}(\mathbf{x}(n), \pmb{\theta}_{l})\bigg{)}, 
\end{equation}
where $\mathcal{L}(\mathbf{x}(n), \pmb{\theta}_{l})$ is the loss function used by the adversary to craft the perturbation and the model that is used to calculate the loss function's gradient is parameterized by $ \pmb{\theta}_{l}$. Thus, the adversarial sample generated using the FGSM targeted at cell $l$ is given by 
\begin{equation} \label{fgsm}
    \tilde{\mathbf{x}}(n) = {\mathbf{x}}(n) + \epsilon \hspace{0.5mm} \text{sign}\bigg{(}\nabla_{\mathbf{x}(n)}\mathcal{L}(\mathbf{x}(n), \pmb{\theta}_{l})\bigg{)}.
\end{equation}
Intuitively, the FGSM aims to add the $\epsilon$-bounded perturbation in the direction that results in the highest loss on the model. This can be considered a step of gradient ascent, where after calculating the direction of the gradient, the perturbation is added in that direction instead of subtracted as is done when training the model using SGD in (\ref{sgd}). This attack is calculated for each DNN deployed at the each of the separate BSs since the attacks considered here are cell-wise. 

\subsection{Projected Gradient Descent (PGD)}
\label{sec:pgd}

PGD is an iterative extension of FGSM. Specifically, instead of adding $\epsilon$ in a single step to each feature in the direction of the loss function's gradient, as in (\ref{delta_fgsm}), a smaller perturbation, $\alpha = \epsilon / Q$, is added over $Q$ iterations, where the direction of the gradient is recalculated on each iteration. Formally, the total perturbation is initialized to zero on the $n^{\text{th}}$ sample, $\mathbf{\Delta}^{(0)}(n) = \mathbf{0}$, and on each iteration, $q$, the perturbation is crafted according to
\begin{equation} \label{pgd_iter}
    \mathbf{\Delta}^{(q)}(n) = \mathbf{\Delta}^{(q-1)}(n) + \alpha \hspace{0.5mm} \text{sign}\bigg{(}\nabla_{\mathbf{x}(n)}\mathcal{L}(\mathbf{x}^{(q-1)}(n), \pmb{\theta}_{l})\bigg{)}, 
\end{equation}
where the final perturbation is given by 
\begin{equation} \label{pgd_delta}
    \pmb{\delta}(n) = \mathbf{\Delta}^{(Q)}(n),
\end{equation}
thus resulting in the $n^{\text{th}}$ adversarial sample crafted according to PGD being 
\begin{equation} \label{pgd_final}
    \tilde{\mathbf{x}}(n) = \mathbf{x}(n) + \mathbf{\Delta}^{(Q)}(n). 
\end{equation}
The increased potency of the PGD attack, over the FGSM, is a result of its iterative recalculation of the optimal direction of attack. This not only makes PGD more potent of an attack, but it also makes it much harder to mitigate at the BS. 

\subsection{Threat Model and Adversary Knowledge Level}

As shown in both FGSM and PGD formulations, the generation of adversarial attacks requires the adversary to calculate the gradient of the loss function with respect to a particular model. However, a successful adversarial attack (i.e., an attack that induces an erroneous power solution by the DNN) does not need to be crafted using the gradient of the DNN under attack. This is because adversarial attacks are transferable \cite{transfer1,transfer2,transfer3} in that attacks crafted to fool one specific DNN are effective on disparate models trained to perform the same task. Due to the transferability property of adversarial attacks, and considering that adversaries will realistically not have precise information about the DNN at the BS \cite{adv_mimo3}, we consider two distinct knowledge levels, termed \emph{threat models}, of the adversary.

Each threat model differs in the amount of system knowledge available to the adversary. The semi-white box threat model assumes that the adversary is aware of the trained DNN model and its parameters but is blind to the defense. The black box threat model assumes that the adversary is blind to both the defense and power prediction DNN. In a white box threat model, by contrast, the adversary would have full knowledge of the DAE and the power allocation DNN as well as their parameters. However, in the context of wireless communications, it is rare for the adversary to have exact knowledge of each signal processing module deployed at the BS \cite{adv_mimo3}. The semi-white box and black box threat models, on the other hand, are more practical to consider because (i) the semi-white box threat model assumes that the adversary has knowledge of the underlying DNN, which may be possible if an off-the-shelf network has been deployed (e.g., residual DNNs for ImageNet \cite{resnet}) and (ii) the black box threat model assumes the adversary does not have access to the defense or the DNN model, which is the most likely attack scenario given that an adversary will rarely have precise information about the signal processing modules at any BS. Thus, due to the impractical nature of white box attacks in massive MIMO networks, we exclude white box attacks from consideration and, instead, show that our defense is robust in semi-white box and black box threat models, which are more pragmatic in real-world maMIMO environments. 

\textbf{Semi-white box attack}: In the semi-white box threat model, we assume that the adversary has full access to the DNN model architecture and its parameters but is blind to any pre-processing functions and algorithms applied to the UE positions prior to its input to the DNN. This type of a threat model would hold when the DNN applied at the BS is an off-the-shelf model made publicly available (as several DNN models are, e.g., residual DNNs for ImageNet \cite{resnet}) giving an adversary access to it but simultaneously leaving them blind to other system modules. Therefore, under the semi-white box threat model, the adversary would use the gradient of the model deployed at the BS to craft the attack. This attack would directly target the model deployed at the BS since its gradient would be used to craft the attacks shown in (\ref{fgsm}) and (\ref{pgd_final}). The semi-white box attack results in a more potent attack in comparison to the black box attack. Higher attack potency, in this context, refers to requiring a smaller perturbation magnitude (i.e., smaller UE positional displacement) to induce an infeasible power allocation prediction from the BS.  

\textbf{Black box attack}: In the black box threat model, the adversary is aware of the power allocation regression task at the BS, but they are blind to both the DNN (including its architecture and parameter values) and the pre-processing applied to the input prior to forward propagating it through the DNN. Here, the adversary leverages the transferability property of adversarial attacks and trains their own surrogate model. Then, the gradient of the surrogate model is used to craft and transfer the attack to the input processed by the DNN at the BS. Specifically, the adversary constructs their surrogate model by first obtaining samples that can be used to train a DNN to perform power allocation prediction. Then, the adversary trains their own DNN architecture, which is different than the architecture deployed at the BS (since the adversary does not have access to the DNN at the BS in the black box threat model). After the adversary's DNN model has been trained, the adversary crafts an adversarial attack according to either (\ref{fgsm}) or (\ref{pgd_final}) using the gradient of their surrogate model. The crafted attack in then injected into samples before they are processed by the DNN at the BS. Due to the transferability property of adversarial attacks, an attack crafted on the adversary's surrogate model is expected to induce an infeasible power allocation solution on the model at the BS with high probability, albeit with lower potency (i.e., the same perturbation displacement may not be as effective in a black box attack compared to a semi-white box attack). 


\section{The Denoising Autoencoder Defense} \label{dae_sec}


\begin{figure}[t] 
	\centering
	\includegraphics[width=\halfwidth\columnwidth]{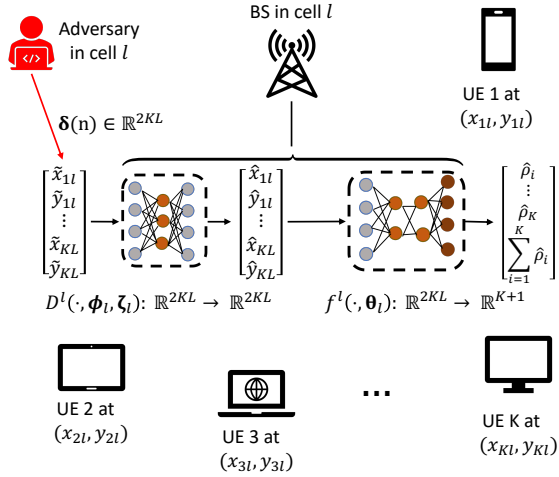}
	\caption{Our considered system diagram within a cell, where the BS contains both the regression DNN and the pre-processing DAE used to mitigate adversarial attacks on UE positions.}
	\label{sys_diagram}
\end{figure}

\begin{table} [t]
\small
\caption{The DAE architecture at each BS \label{dae_arch}}
\centering
\begin{tabular}{c | c | c } 
\centering 
\makecell{Layer} & \makecell{Units } & \makecell{Activation \\ Function } \\ 
\hline
Input & $KL$ & - \\
Layer 2 & 32 & Linear \\
Layer 3 & 16 & Linear \\
Layer 4 & 32 & Linear \\
Output & $KL$ & Linear \\

\hline

\end{tabular}
\end{table}

In this section, we discuss our proposed methodology for defending maMIMO power allocation DNNs from adversarial attacks. We begin by introducing a deep learning-based denoising function, which is used to pre-process inputs prior to their input to the DNN. After the DAE has been trained, samples at inference time (when the DAE is deployed at the BS) are forward propagated through the DAE before being inputted into the regression DNN. The DAE is trained to mitigate the effect of any potential additive adversarial perturbation as well as retain performance on unperturbed signals. 

\begin{algorithm}[t] 
   \caption{Denoising Autoencoder Construction}
   \label{dae_const}
   \begin{algorithmic}[1] 
        \STATE \textbf{input:} $\mathcal{X}_{\text{tr}}$: Set of input samples: $\mathbf{x}(n) \hspace{2mm} \forall \hspace{2mm} n$ \\ 
        \hspace{9mm} $f^{l}(\cdot, \pmb{\theta}_{l})$: Trained regression DNN at BS $l$ \\
        \hspace{9mm} $\alpha_{l}$: $l_{\infty}$-norm bound used at BS $l$ \\
        \hspace{9mm} $u^{l}(\cdot, \pmb{\phi}_{l})$: Randomly initialized encoder function \\
        \hspace{9mm} $v^{l}(\cdot, \pmb{\zeta}_{l})$: Randomly initialized decoder function \\
        
        \STATE \textbf{initialize:} $\mathcal{X}_{\text{DAE}} \gets \emptyset$ \\
        
        \FOR{$\mathbf{x}(n) \in \mathcal{X}_{\text{tr}}$} 
        
            \STATE $\mathbf{\Delta}^{(0)}(n) = \mathbf{0}$
        
            \FOR{$q = 1, \ldots, Q$} 
                
                \STATE $\mathcal{L}(\mathbf{x}^{(q-1)}(n), \pmb{\theta}) = \sum_{k=1}^{K} \hat{\rho}_{jk}$
                \STATE $\pmb{\kappa}^{(q-1)}(n) = \text{sign}(\nabla_{\mathbf{x}(n)}\mathcal{L}(\mathbf{x}^{(q-1)}(n), \pmb{\theta}_{l}))$
                \STATE $\mathbf{\Delta}^{(q)}(n) = \mathbf{\Delta}^{(q-1)}(n) + \alpha_{l} \hspace{0.5mm} \pmb{\kappa}^{(q-1)}(n)$ 
        
            
            \ENDFOR 
            
            \STATE $\tilde{\mathbf{x}}(n) = \mathbf{x}(n) + \mathbf{\Delta}^{(Q)}(n)$
            \STATE append $\tilde{\mathbf{x}}(n)$ to $\mathcal{X}_{\text{DAE}}$
        
        \ENDFOR 
        \STATE $\mathcal{X}_{\text{DAE}} \gets \text{concatenate}(\mathcal{X}_{\text{DAE}}, \mathcal{X}_{\text{tr}})$
        \STATE $\mathcal{X}_{\text{target}} \gets \text{concatenate}(\mathcal{X}_{\text{tr}}, \mathcal{X}_{\text{tr}})$
        
        \FOR{$\mathbf{x}'(n), \mathbf{x}(n) \in \mathcal{X}_{\text{DAE}}, \mathcal{X}_{\text{target}}$} 
        
            \STATE $\mathbf{e} = \mathbf{x}(n) - v^{l}(u^{l}(\mathbf{x}'(n)))$
            \STATE $D^{l}(\cdot, \pmb{\phi}_{l}, \pmb{\zeta}_{l}) \gets \underset{\pmb{\phi}_{l}, \pmb{\zeta}_{l}} {\text{min}} \frac{1}{2KL} \sum_{s=1}^{2KL} (\mathbf{e}(s))^2$
        
        
        \ENDFOR 

        \RETURN $D^{l}(\cdot, \pmb{\phi}_{l}, \pmb{\zeta}_{l})$
  
  \end{algorithmic}

\end{algorithm}

Adversarial attacks introduce subtle perturbations into the input that lead to erroneous outputs by the DNN. Since such attacks are often gradient-based, they will be added in a manner consistent with maximizing some loss function regardless of the type of attack and threat model adopted by the adversary. We leverage this intuition and propose training a Denoising Autoencoder (DAE), which is capable of filtering noise from a consistent corruption processes \cite{dl_book}. 



The construction of our proposed DAE is detailed in Algorithm \ref{dae_const}. We begin, in lines 3 -- 12, by using the samples comprising $\mathcal{X}_{\text{tr}}$ to generate adversarial examples using PGD according to (\ref{pgd_iter}) -- (\ref{pgd_final}). Our DAE is specifically tailored to defend PGD because, as a result of its iterative recalculation of the gradient as discussed in Sec. \ref{sec:pgd}, adversarial examples generated using PGD capture an exhaustive set of gradient-based adversarial attacks. Therefore, as we show in Sec. \ref{results_sec}, defenses trained against PGD attacks effectively generalize to mitigate other classes of gradient-based attacks such as FGSM. Furthermore, prior work \cite{pgd} has shown that, for classification-based DNNs, robustness against this attack guarantees robustness against other first-order $l_{\infty}$-bounded adversarial attacks as well. In Sec. \ref{results_sec}, we show that this property extends to regression-based DNNs as well in the specific context of mitigating adversaries in maMIMO power allocation. 


In the implementation of Algorithm \ref{dae_const}, PGD perturbations are generated using the gradient of the DNN at BS $l$. This corresponds to a semi-white box adversarial sample generation method, which is chosen since semi-white box attacks are more potent than black box attacks. Thus, training to defend a semi-white box attack is expected to increase robustness to black box attacks, despite the gradient used to generate such attacks being in a suboptimal direction of attack (different than the direction of the gradient of the loss calculated using the DNN under attack at the BS).

Following the generation of PGD adversarial examples, at BS $l$, we train a DAE denoted by $D^{l}(\cdot, \pmb{\phi}_{l}, \pmb{\zeta}_{l}): \mathbb{R}^{2KL} \rightarrow \mathbb{R}^{2KL}$, which consists of an encoder function, $u^{l}(\cdot, \pmb{\phi}_{l}): \mathbb{R}^{2KL} \rightarrow \mathbb{R}^{z}$, parameterized by $\pmb{\phi}_{l}$, and a decoder function $v^{l}(\cdot, \pmb{\zeta}_{l}): \mathbb{R}^{z} \rightarrow \mathbb{R}^{2KL}$ parameterized by $\pmb{\zeta}_{l}$. The encoder compresses the $2KL$ dimensional input into a $z$ dimensional representation, and the decoder reconstructs an approximation of the uncorrupted input from the $z$ dimensional representation to the original $2KL$ dimensional space. The full DAE, comprised of both the encoder and decoder, are constructed using fully connected layers. The architecture of our proposed model is shown in Table \ref{dae_arch}. After random initialization, the DAE is trained, using SGD, to map the generated PGD adversarial examples to an approximation of the unperturbed positional input by minimizing the mean squared error. Here, we first calculate the error between the DAE's reconstruction of an adversarial example, and its corresponding clean sample given by
\begin{equation} \label{dae_error}
    \mathbf{e} = \mathbf{x}(n) - v^{l}(u^{l}(\tilde{\mathbf{x}}(n))),
\end{equation}
where $v^{l}(u^{l}(\tilde{\mathbf{x}}(n)))$ is the output of the DAE on input $\tilde{\mathbf{x}}(n)$ and $\mathbf{e} \in \mathbb{R}^{2KL}$, which corresponds to the difference being taken element-wise. Next, we train the DAE by minimizing the squared error given by
\begin{equation} \label{dae_mse}
    \underset{\pmb{\phi}_{l}, \pmb{\zeta}_{l}} {\text{min}} \quad \frac{1}{2KL} \sum_{s=1}^{2KL} (\mathbf{e}(s))^2, 
\end{equation}
where $\mathbf{e}(s)$ denotes the $s^{\text{th}}$ element in the error vector $\mathbf{e}$. Although the formulation shown here corresponds to training on one sample at a time (i.e., a batch size of 1), mini-batches consisting of multiple samples are used in practice to accelerate training. 

During the deployment of the DAE defense, the BS will be unaware whether the input is corrupted or not (i.e., the BS will not know whether $\mathbf{x}(n)$ of $\tilde{\mathbf{x}}(n)$ was received). Therefore, while the trained DAE should map $\tilde{\mathbf{x}}(n)$ to an approximation of $\mathbf{x}(n)$, it should also be capable of mapping $\mathbf{x}(n)$ to an approximation of itself so that benign inputs do not suffer performance degradation by the defense. During training, we enforce this mapping by including all uncorrupted training samples (i.e., all samples in $\mathcal{X}_{\text{tr}}$) in the training set for the DAE. This is shown in lines 13 -- 18 of Algorithm \ref{dae_const}. 

After training the DAE, each input is pre-processed using $D^{l}(\cdot, \pmb{\phi}_{l}, \pmb{\zeta}_{l})$ before being inputted to the DNN for power allocation. Specifically, given a received sample, $\tilde{\mathbf{x}}(n)$, we first denoise it to subtract any potential adversarial perturbation. The denoised input is given by $\hat{\mathbf{x}}(n) = D^{l}(\tilde{\mathbf{x}}(n), \pmb{\phi}_{l}, \pmb{\zeta}_{l})$. Then, $\hat{\mathbf{x}}(n)$ is forward propagated through the DNN regression model to yield the prediction power allocation solution to each UE in cell $j$ given by $\hat{\pmb{\rho}}_{j}(n) = f^{l}(\hat{\mathbf{x}}(n), \pmb{\theta}_{l})$. Our overall system diagram is shown in Fig. \ref{sys_diagram}. 

The threat model under which the adversary operates determines the type and potency of the launched attack. In the semi-white box threat model, the adversary uses the gradient of the model at the BS to craft its attack using either (\ref{fgsm}) or (\ref{pgd_final}). In this threat model, the adversary is blind to the DAE parameters and processing function at the BS. Moreover, this threat model results in the most potent attack since it directly targets the DNN employed at the BS. Contrarily, in the black box threat model, the adversary is blind to both the DAE and the DNN employed at the BS. As a result, the adversary instantiates an attack using the gradient of a surrogate model (i.e., a model that the adversary obtained access to by training it to perform the same task as the DNN at the BS) and relies on the transferability property of adversarial attacks to induce infeasible power allocation predictions at the BS. Although this attack is less potent than the semi-white box attack, it still reduces the DNN's performance and is the most likely attack that the adversary will induce, given that adversaries will typically be knowledge-limited about the signal processing modules at the BS. 

Finally, we analyze the online computational complexity of our proposed DAE defense. For a general DNN, a single matrix followed by an element-wise activation function is needed to characterize the weights and transformations between one layer to the next. Assuming an $A \times B$ weights matrix with $A > B$, where $A$ and $B$ are the number of units in the former and latter layers, respectively, the computational complexity of forward propagating a sample through a single layer with an activation function is given by $\mathcal{O}(A^{3})$. Thus, the computational complexity for a $p$ layered DNN is upper bounded by $\mathcal{O}(pA^{3})$ (assuming that the largest number of units in any layer of the network is $A$). Our DAE introduces an additional complexity of $\mathcal{O}(sa^{2})$, where the complexity is squared instead of cubed because the activation function is linear, $a < A$ since the largest layer of the DAE has less units than the largest layer in the DNN (e.g., as in Tables \ref{model1}, \ref{model2}, and \ref{dae_arch}), and $s$ is the number of layers in the DAE. Therefore, the total time complexity of our proposed method is given by $\mathcal{O}(sa^{2} + pA^{3})$. This leads to computationally efficient runtimes, as we demonstrate empirically in Sec. \ref{sec:miti_eff}. 


\section{Performance Evaluation} \label{results_sec}

In this section, we conduct an empirical evaluation of our methodology. We describe our empirical setup, show the performance of our defense in both the presence and absence of adversarial attacks, and discuss the computational efficiency of our method.  


\subsection{Dataset and Performance Metrics}

\begin{table} [t]
\small
\caption{The parameter values used to train our proposed defense in our empirical evaluation.} \label{sim_params}
\centering
\begin{tabular}{c | c  } 
\centering 
Parameter & Value \\ 
\hline
$L$ & $4$ cells \\
$M$ & $100$ antennas \\
$K$ & $5$ UEs \\
Cell area & $250 \text{m}$ $\times$ $250 \text{m}$ \\
$P_{\text{max}}^{\text{dl}}$ & $500$ mW\\
$\sigma^2$ & $-94$ dBm\\
Bandwidth & $20 \text{MHz}$ \\
$z$ & $16$ units \\
$\alpha_{l}$ & $0.03$ \\
$Q$ & $10$ iterations \\
$N$ & $329000$ samples \\ 
$B$ & $500$ samples \\

\hline

\end{tabular}
\end{table}



To assess the efficacy of our methodology, we employ the Power Allocation in Multi-Cell Massive MIMO dataset, which is publicly available at \url{https://data.ieeemlc.org/Ds2Detail}. All parameter values used in our evaluation are presented in Table \ref{sim_params}. We consider $L = 4$ cells in a square grid layout of $2 \times 2$ cells. In each cell, the BS is located in the center of the cell. The presented results are averaged across all BSs. A feasible power allocation solution is achieved when the DNN predicts a total power allocation less than or equal to $P_{\text{max}}^{\text{dl}}$ to all UEs in the cell. Otherwise, the solution in infeasible. 

\begin{figure}[t]
  \centering
  \includegraphics[width=\halfwidth\columnwidth]{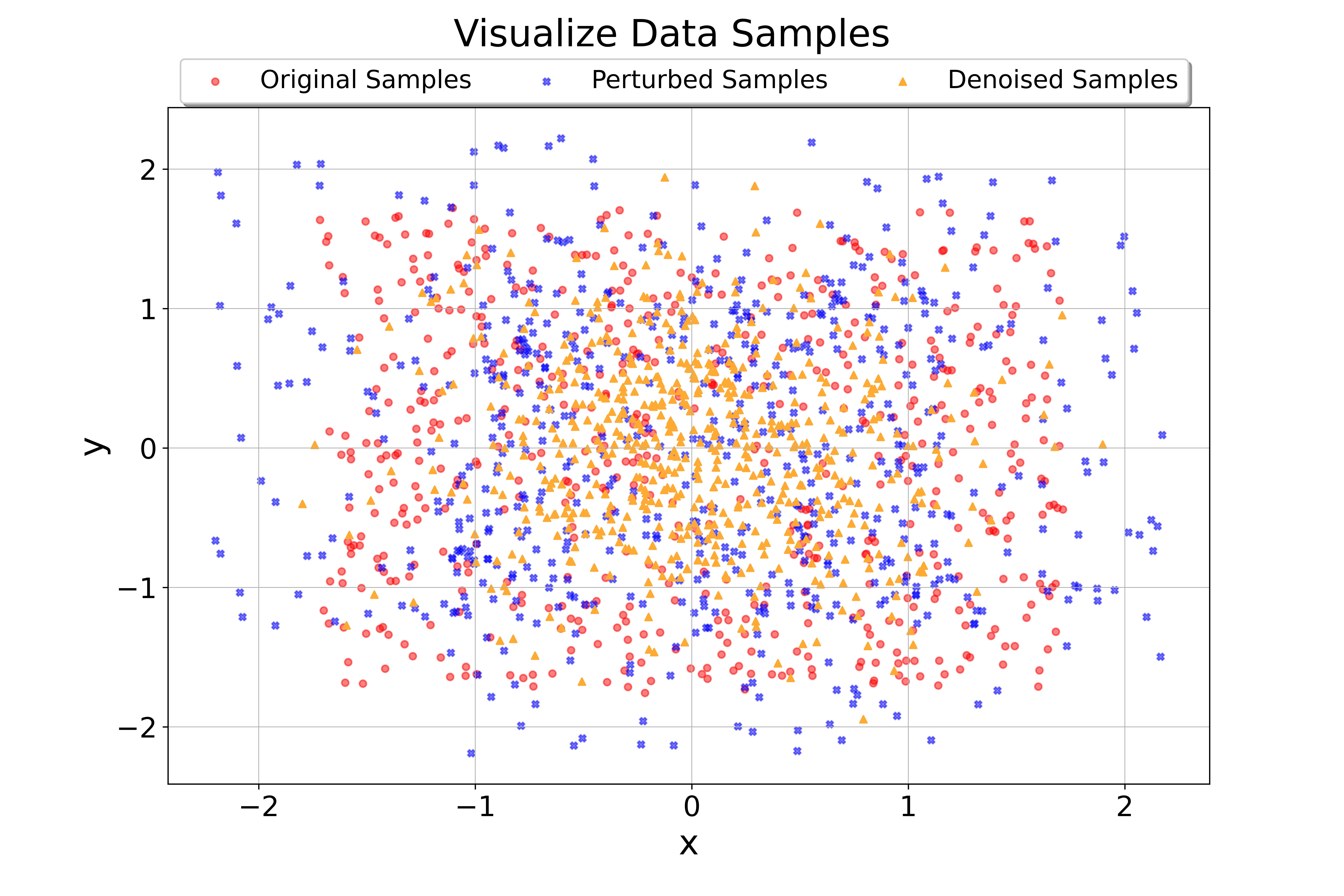}
  \caption{40 testing samples shown in their original unperturbed form, perturbed form, and denoised form after forward propagating through the DAE. Each point denotes the position of the UE. We see that perturbed samples deviate substantially from their unperturbed counterparts, whereas denoised samples are much closer to unperturbed samples.}
  \label{samples_pltd}
\end{figure}

The DAE is trained according to Algorithm \ref{dae_const} using the architecture and parameter values shown in Tables \ref{dae_arch} and \ref{sim_params}, respectively. Fig. \ref{samples_pltd} visually compares some examples of the original, adversarial, and denoised samples. At test time, we vary the perturbation magnitude of each considered attack and measure the \emph{attack success rate}, which is the number of samples in the testing set that exceed the power budget of $P_{\text{max}}^{\text{dl}} = 500 \text{mW}$ in the DNN prediction output (i.e., the proportion of samples that satisfy each constraint in (\ref{adv_opt:all-lines})). The inputs to each model (Model 1, Model 2, and DAE) are the normalized UE positions with zero mean and unit variance (corresponding to uniformly distributed UEs) and, we consider attacks crafted directly on the normalized samples. As a result, the perturbation magnitude induced on each normalized input sample corresponds to a physical UE displacement of $\sqrt{2}\psi\epsilon$, where $\psi$ is the standard deviation of the unnormalized samples. Here, scaling by a factor of $\psi$ unnormalizes the sample's positional scale, and the factor of $\sqrt{2}$ accounts for the total UE displacement since each UE is displaced along both geographical coordinates by $\epsilon$. Note that the standard deviation used to normalize each feature is $\psi \approx 145$, which corresponds to the standard deviation of a uniform distribution with a $500$ m interval (i.e., the assumed UE distribution in the $2 \times 2$ cells of $250 \times 250$ m each considered here). For example, a perturbation magnitude of $\epsilon = 0.05$ in the FGSM attack corresponds to a maximum UE displacement of $d_{\epsilon} = 10.28$ m. Similarly, for the PGD attack with $Q = 10$ iterations, a perturbation magnitude of $\alpha = 0.01$ corresponds to a a maximum UE displacement, as observed by the BS, of $d_{\epsilon} = 20.57$ m.



\subsection{Semi-white Box Performance}  

\begin{figure}[t]
  \centering
  \includegraphics[width=\halfwidth\columnwidth]{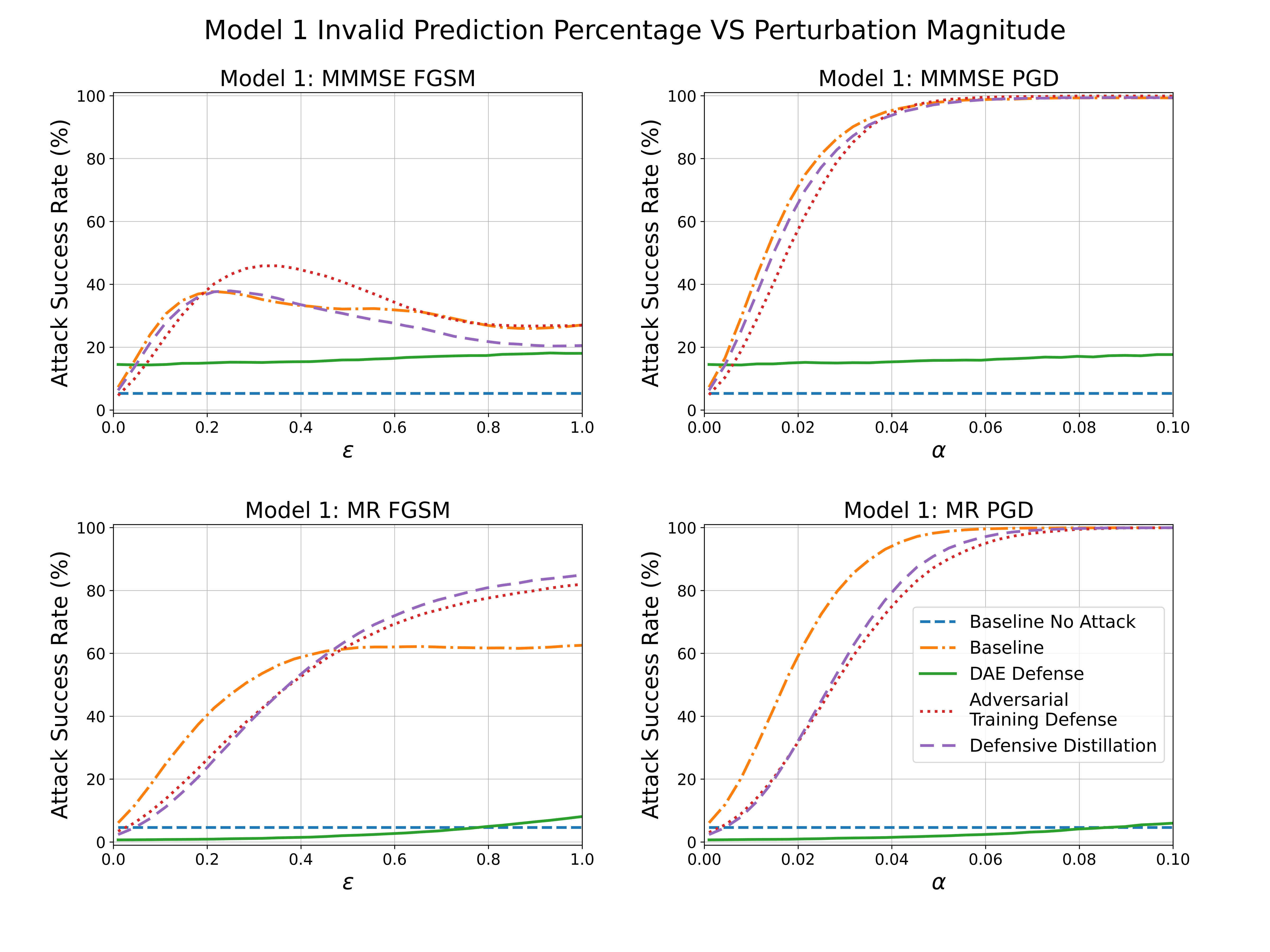}
  \caption{Semi-white box attack on Model 1 for each considered attack and precoding scheme. We see that, on undefended models, the attack success rate grows as the perturbation magnitude increases. Our proposed DAE results in a significantly lower attack success rate outperforming the considered baselines.}
  \label{whitebox_m1}
\end{figure}

\begin{figure}[t]
  \centering
  \includegraphics[width=\halfwidth\columnwidth]{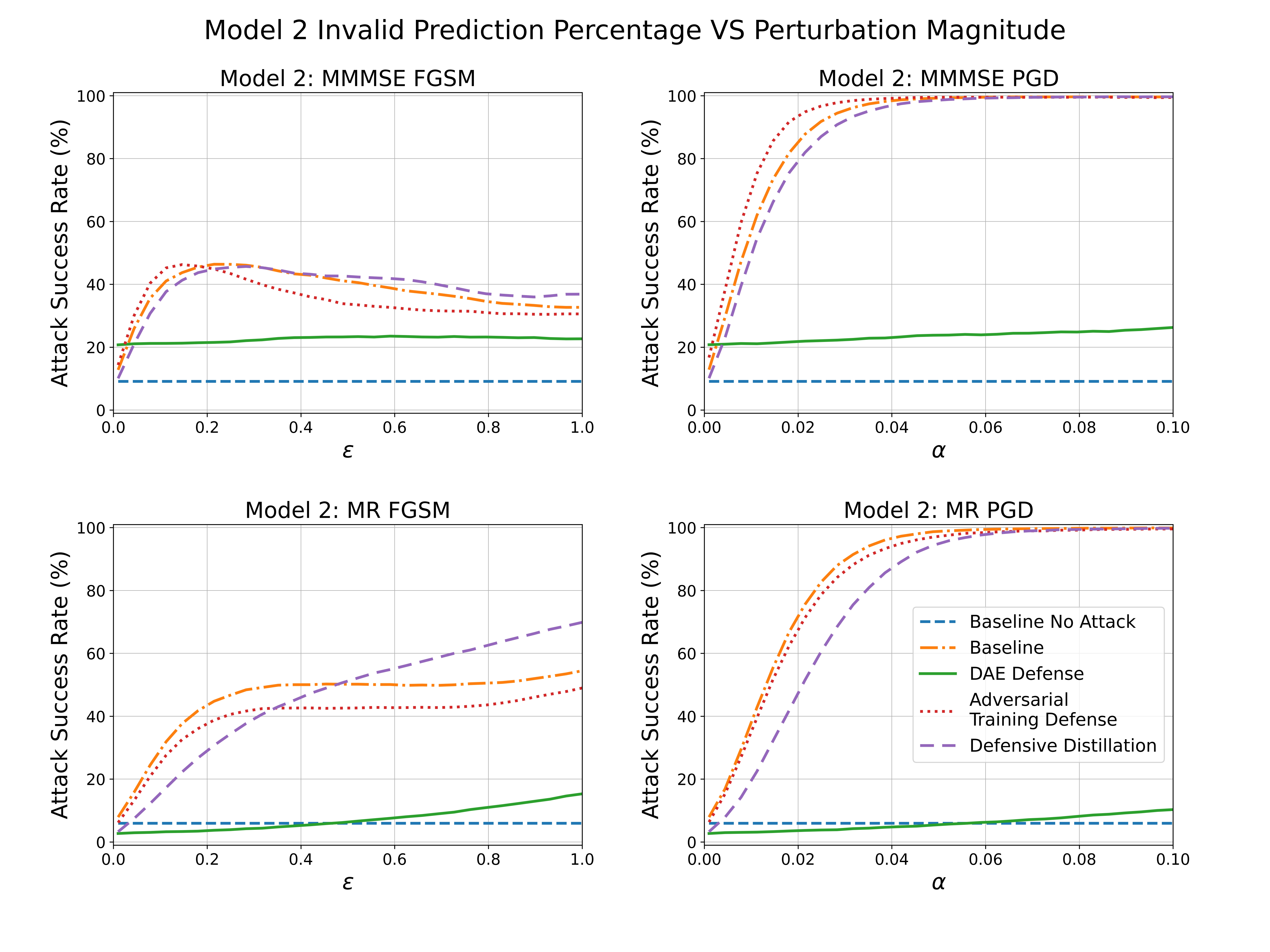}
  \caption{Semi-white box attack on Model 2 for each considered attack and precoding scheme. Similar to Fig. \ref{whitebox_m1}, we see that the PGD attack is more potent than the FGSM attack. In addition, we see that our proposed DAE outperforms the considered baseline and significantly reduces the attack success rate in each considered case.}
  \label{whitebox_m2}
\end{figure}


\begin{figure}[t]
  \centering
  \includegraphics[width=\halfwidth\columnwidth]{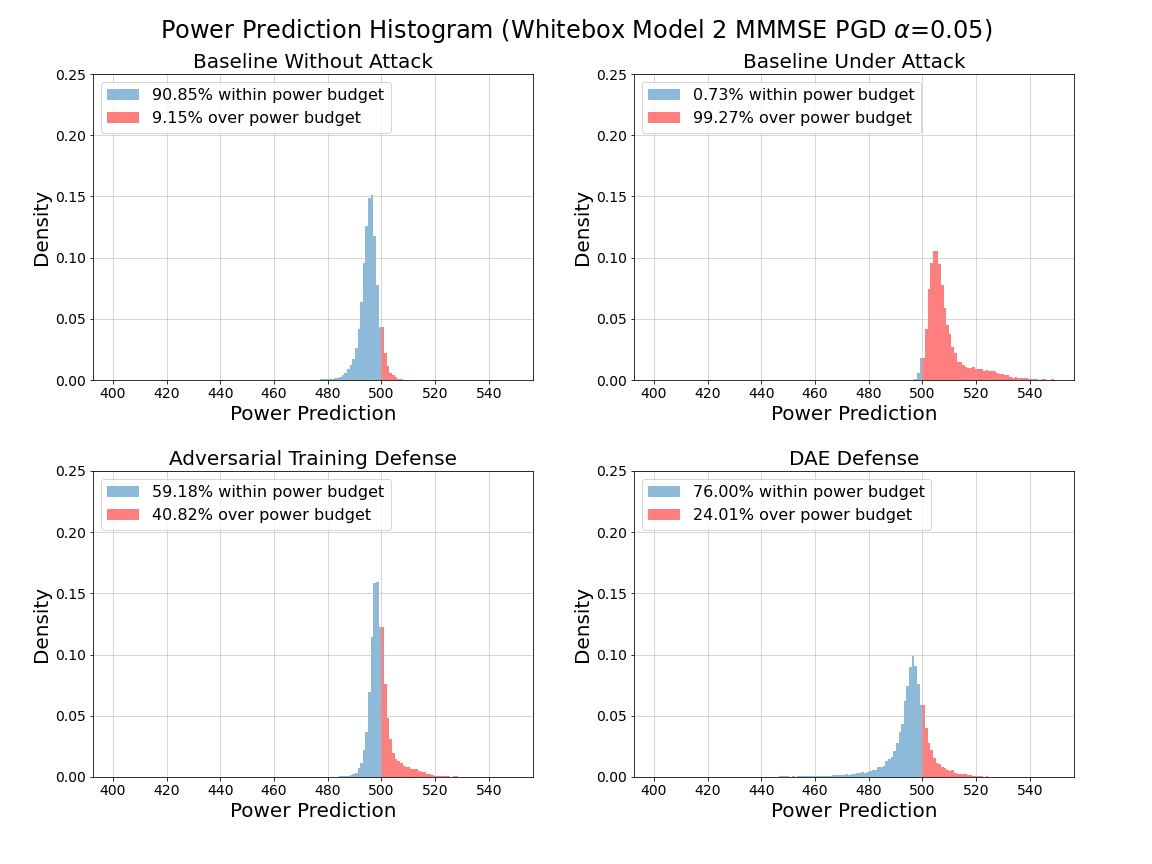}
  \caption{The power prediction distributions under each considered defense scenario on Model 2 in the semi-white box environment. Wee see that our proposed DAE defense (bottom right) significantly lowers the proportion of samples predicted with infeasible power solutions (i.e., when $P_{\text{max}}^{\text{dl}} > 500 \text{mW}$) over using no defense (top right) and adversarial training (bottom left).}
  \label{hist_whitebox}
\end{figure}

We begin by evaluating the efficacy of our proposed defense in the semi-white box threat model, where the adversary has access to the classifier (either Model 1 or Model 2) under attack and uses its gradient to generate an adversarial perturbation. We consider attacks on both Model 1 and Model 2 using both the MMMSE and MR precoding schemes. In addition, we compare our results to adversarial training on maMIMO DNN models \cite{mimo_adv_trn} as well as defensive distillation \cite{def_dist} in each case as a performance benchmark. Adversarial training, as proposed in \cite{mimo_adv_trn}, consists of generating adversarial perturbations on the target DNN (i.e., the model whose gradient will be used to generate the attack) and retraining it with the corrected labels. Although this method is effective in mitigating low-bounded perturbations, it remains vulnerable to higher bounded perturbations and, furthermore, is highly susceptible to overfitting on adversarial samples thus lowering its prediction performance on unperturbed inputs \cite{adv_trn_ovft}. Defensive distillation first trains each DNN model on the true ``hard labels" (i.e., the true power allocation vectors). Then, training samples are forward propagated through the DAE to obtain each samples' ``soft labels," which are the predicted labels from the trained model. The original input samples and the ``soft labels" are then used to train a new model, which serves as the defense against adversarial attacks. Defensive distillation applied for regression tasks, however, may potentially hinder baseline performance, since it forces re-training of the model on inaccurate power allocation outputs (i.e., the ``soft labels"). Our proposed method aims to address these shortcomings of both adversarial training and defensive distillation by avoiding model retraining. 

To evaluate and compare our proposed method, we show four performance curves per environment. First, we show the baseline with no attack, which is the proportion of samples whose power predictions naturally result in infeasible power solutions without any added perturbations. This is analogous, in a classification setting, to samples that are misclassified due to the DNN's inability to correctly predict them rather than due to any additive perturbations. Thus, since the baseline is not dependent on the perturbation magnitude, it stays constant throughout the considered perturbation range. Second, we show the performance curves for the baseline case, which is the performance of the DNN without any defense in place to mitigate the attack. Note that the attack success rate for the baseline case, in the majority of settings, increases with higher perturbation magnitudes thus confirming that the approximated attack solution is not guaranteed to effectively induce an infeasible power allocation solution from the DNN. Third, we show the performance of our DAE defense, and lastly, we show the performance of adversarial training and defensive distillation as previously described. 


Figs. \ref{whitebox_m1} and \ref{whitebox_m2} show the DNN performance of Model 1 and Model 2, respectively. In general, we can see from both models that the PGD attack is more potent than the FGSM attack irrespective of the precoding scheme. In addition, we see that both adversarial training and defensive distillation remain susceptible to high-bounded adversarial attacks which is consistent with prior works that have considered using them as a defense \cite{mimo_adv_trn,adv_trn_ovft,def_dist_not_rob}. This trend is observed under both FGSM and PGD perturbations, where adversarial training and defensive distillation deliver slightly improved performance over no defense. Our proposed defense, on the other hand, is able to significantly mitigate semi-white box attacks across the entire range of considered perturbation magnitudes. For example, at a perturbation magnitude of $\alpha = 0.10$, the PGD attack on MR pre-coded samples on both Models 1 and 2 results in nearly $100\%$ samples being predicted with infeasible power allocation solutions on undefended models, models with adversarial training, and defensive distillation. However, with our proposed DAE in place, only approximately $10\%$ of the samples produce infeasible solutions. Although the FGSM attack yields a smaller proportion of infeasible solutions in comparison to the PGD perturbation, our defense is still able to mitigate its effects to a large degree (e.g., from $60\%$ to $10\%$ on Model 1 with MR precoding). These observations are consistent across both Model 1 and Model 2 in defending both considered attacks on both considered precoding schemes. 

\begin{table} [t]
\small
\caption{Adversary's attack success rate for each model on unperturbed inputs. For MR precoded signals, we see that the DAE defense improves performance on perturbed signals. On MMMSE precoded signals, the baseline performance experiences a slight tradeoff for significant performance improvements against high-bounded attacks (further shown in Figs. \ref{whitebox_m1}, \ref{whitebox_m2}, \ref{blackbox_m1}, and \ref{blackbox_m2}).} \label{clean_res}
\centering
\begin{tabular}{c | c | c | c | c} 
\centering 
 & \makecell{Model 1 \\ MMMSE} & \makecell{Model 1 \\ MR } & \makecell{Model 2 \\ MMMSE} & \makecell{Model 2 \\ MR} \\ 
\hline
Baseline DNN & 5.31\% & 4.63\%  & 9.15\% & 5.97\% \\
DAE Defense & 14.57\% & 0.67\% & 20.64\% & 2.69\% \\

\hline

\end{tabular}
\end{table}

Next, we consider the performance of our defense in the absence of adversarial attacks. Note that, upon receiving a sample of UE positional locations, the BS will be unaware of whether it contains an adversarial perturbation or not. Therefore, any employed defense should keep the baseline performance on unperturbed signal intact. In Figs. \ref{whitebox_m1} and \ref{whitebox_m2}, unperturbed signals correspond to $\epsilon = 0$ and $\alpha = 0$ for the FGSM and PGD perturbations, respectively (this results in $\pmb{\delta}(n) = \mathbf{0}$ in (\ref{delta_fgsm}) and (\ref{pgd_delta}) for FGSM and PGD, respectively). From Figs. \ref{whitebox_m1} and \ref{whitebox_m2}, we see that, at a perturbation magnitude of zero, our proposed defense retains approximately the same performance as the baseline model with no attack (i.e., the performance on the original DNN at the BS before employing any defense). In fact, the DAE slightly improves the baseline power prediction performance in the absence of adversarial attacks in black box setups with MR precoding. The performance of our model in the absence of attacks is shown in Table \ref{clean_res}. Ultimately, in all cases, the proposed defense provides strong attack mitigation abilities on a wide range of perturbation magnitudes. 

We more closely examine the distribution of predicted power by the DNN under different defenses in Fig. \ref{hist_whitebox}. Here, we instantiate a PGD attack using $\alpha = 0.05$ on Model 2. Once again noting that $P_{\text{max}}^{\text{dl}} = 500 \text{mW}$, we see that the original set of unperturbed testing samples only result in $9.15\%$ of samples being predicted with infeasible power solutions whereas, after instantiating the attack with no defense, $99.27\%$ of predicted samples are infeasible. Our proposed DAE defense significantly reduces the proportion of samples predicted with infeasible solutions to $24.01\%$, which outperforms adversarial training and more closely approaches the performance of the model with no attack. 

Finally, as shown in Algorithm \ref{dae_const}, only PGD adversarial examples are generated to train the DAE. However, as shown in Figs. \ref{whitebox_m1} and \ref{whitebox_m2}, the DAE is still able to defend the FGSM perturbation. Thus, we see that training to defend the PGD attack, as our proposed defense does, provides robustness against other first-order $l_{\infty}$-bounded perturbations on power allocation regression-based DNNs in maMIMO. Furthermore, the PGD attack generated to defend the DNN only employs one perturbation bound ($\alpha = 0.03$ and $Q = 10$ as discussed earlier). Yet, it is able to provide mitigation against a wide range of adversarial perturbation magnitudes as demonstrated in Figs. \ref{whitebox_m1} and \ref{whitebox_m2}. This provides a substantial improvement over adversarial training, where training on multiple bounds may be required for better robustness at the cost of lower performance on non-adversarial samples. Training our proposed DAE on a single perturbation bound not only retains the performance on unperturbed samples (i.e., when $\pmb{\delta}(n) = \mathbf{0}$), but it also does not incur costly computational overhead during deployment, which will be further discussed in Sec. \ref{sec:miti_eff}. 

\subsection{Black Box Performance}


\begin{figure}[t]
  \centering
  \includegraphics[width=\halfwidth\columnwidth]{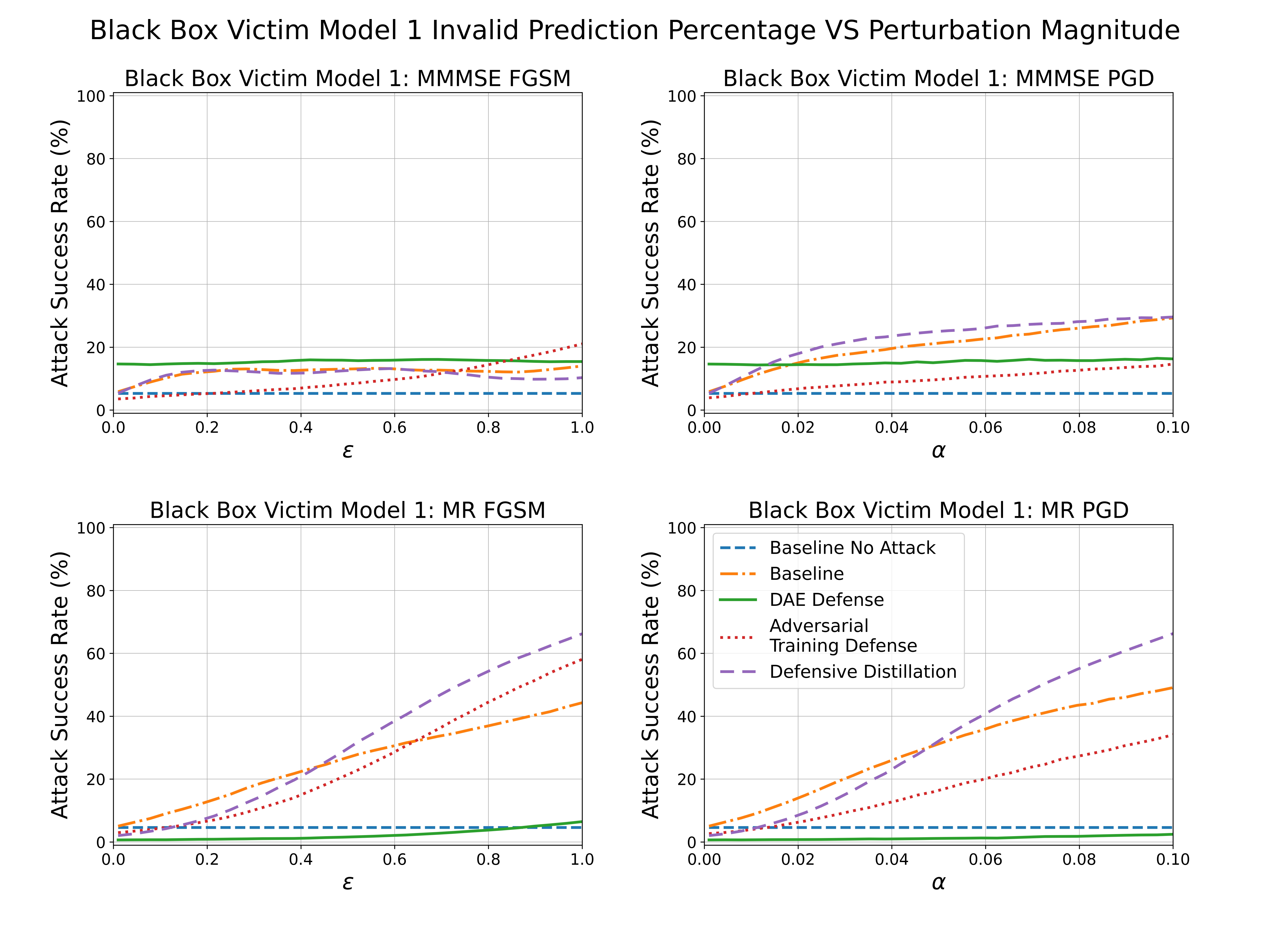}
  \caption{Black box attack where the adversary crafts their attack using Model 2, and Model 1 is the victim model employed at the BS. In comparison to the semi-white box scenario, we see that black box adversarial attacks are less potent on the victim model. Nonetheless, our proposed DAE still reduces the attack success rate over undefended and baseline methods in each considered setup.}
  \label{blackbox_m1}
\end{figure}


\begin{figure}[t]
  \centering
  \includegraphics[width=\halfwidth\columnwidth]{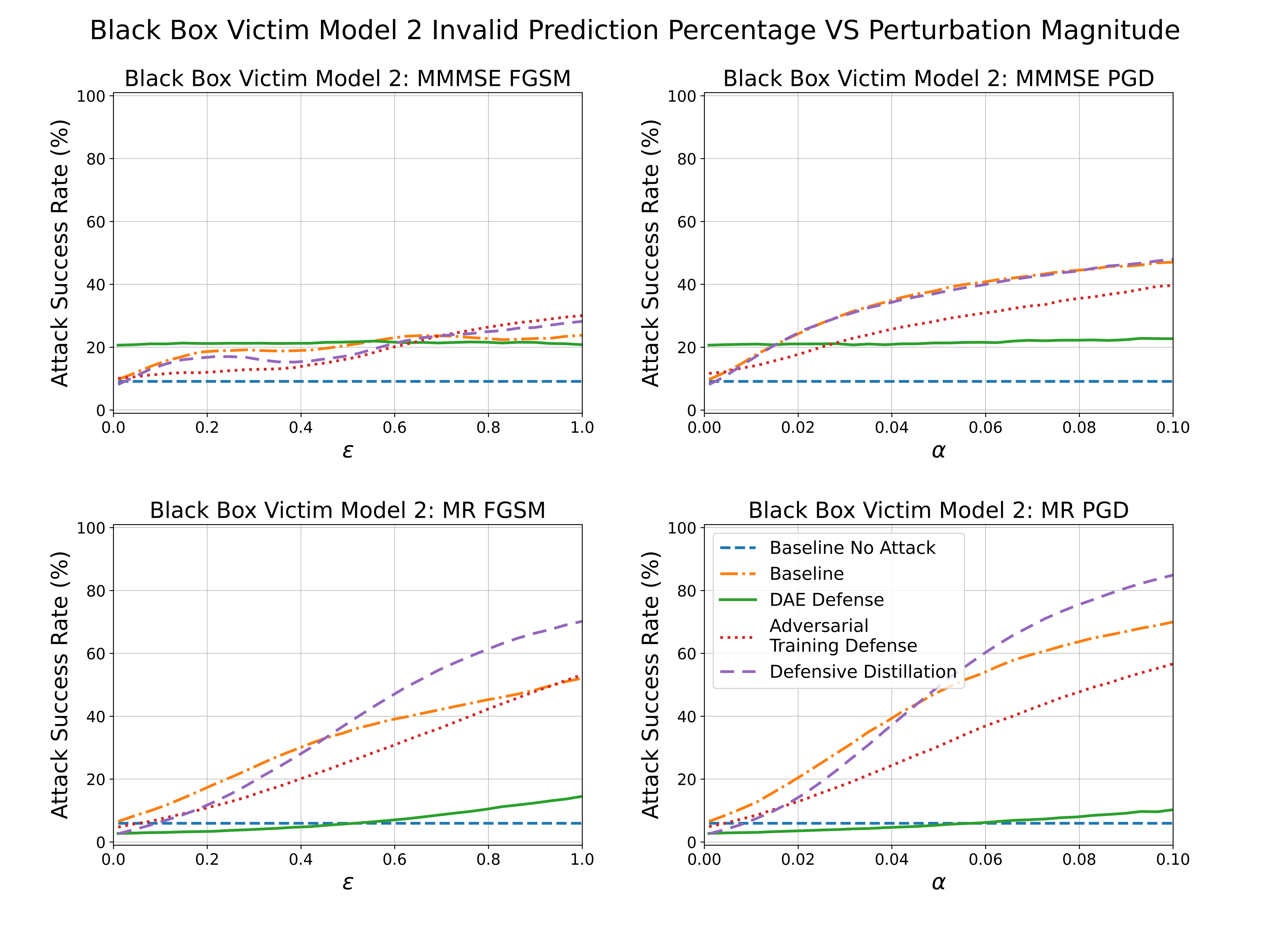}
  \caption{Black box attack where the adversary crafts their attack using Model 1, and Model 2 is the victim model employed at the BS. Similar to Fig. \ref{blackbox_m1}, we see that black box attacks are less potent than semi-white box attacks, but our proposed method is still the most effective in reducing the adversary's success rate in each considered scenario.}
  \label{blackbox_m2}
\end{figure}

\begin{figure}[t]
  \centering
  \includegraphics[width=\halfwidth\columnwidth]{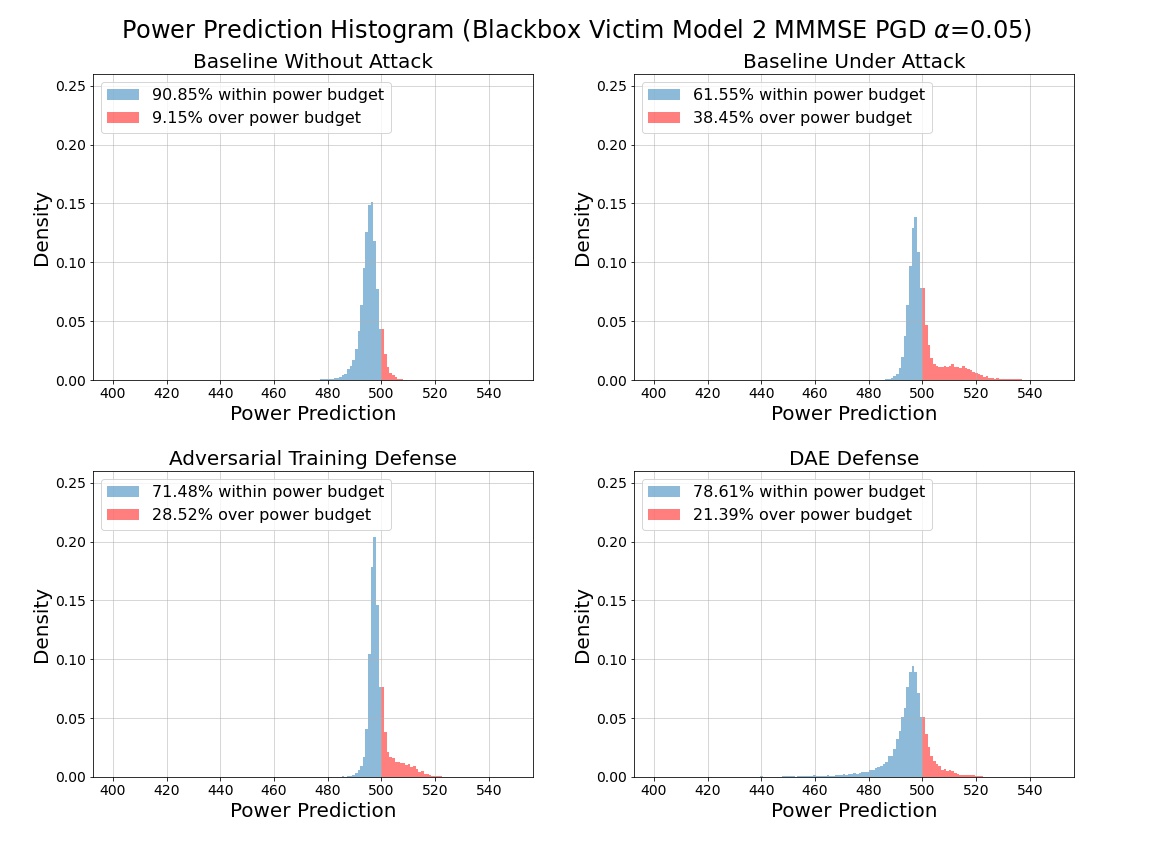}
  \caption{The power prediction distributions under each considered defense scenario on Model 2 in the black box environment. Similar to Fig. \ref{hist_whitebox}, we see that our proposed DAE defense (bottom right) lowers the proportion of samples predicted with infeasible power allocation solutions ($P_{\text{max}}^{\text{dl}} > 500 \text{mW}$), outperforming both undefended models (top right) and adversarial training (bottom left).}
  \label{hist_blackbox}
\end{figure}



Now, we discuss the effectiveness of our defense in the black box threat model, where the adversary is blind to the DNN model used for power allocation at the BS. Here, we consider two attack scenarios. The first is where the adversary crafts an adversarial attack using the gradient of Model 2 but Model 1 is used to predict the power allocation at the BS using the perturbed sample. The second case is where the attack is crafted using the gradient of Model 1 but the perturbed sample is evaluated on Model 2 at the BS. Results for the former and latter attack scenarios are shown in Figs. \ref{blackbox_m1} and \ref{blackbox_m2}, respectively, where we use the term \emph{victim model} to refer to the model under attack and the term \emph{adversary model} to refer to the model whose gradient is used to generate the perturbation.

From Figs. \ref{blackbox_m1} and \ref{blackbox_m2}, we see that black box attacks are generally less potent on regression-based DNNs for power allocation in comparison to semi-white box attacks. This is most likely due to the gradient direction in which the perturbation should be added to induce infeasible solutions on the victim model differing from the adversary's model to a significant degree. We find that this trend, which has been shown in prior work for classification DNNs \cite{transfer1,transfer2}, extends to black box attacks on regression DNNs for power allocation. This is further shown in Fig. \ref{hist_blackbox}, where we plot the distribution of the sum of the predicted power for all testing samples. In comparison to Fig. \ref{hist_whitebox}, corresponding to the semi-white box threat model, we see that the same perturbation magnitude ($\alpha = 0.05$) results in less infeasible samples on defended models, thus corroborating the lower potency of black box attacks in maMIMO DNNs. Despite the lower potency, we still see that our proposed defense, in general, is able to lower the proportion of samples predicted with infeasible power solutions. In certain cases, such as black box attacks induced on MMMSE precoded signals using the FSGM perturbation, the proposed defense slightly improves the attack success rate. Thus, in these cases, the BS experiences a slight performance degradation, which is traded off for significant attack mitigation in all other threat models. 

Moreover, we see that the general trends observed on the effectiveness of our defense in the semi-white box threat model hold in the black box threat model as well despite the attacks being less transferable (and thus less potent in the black box threat model). For example, the DAE retains its prediction performance on unperturbed samples as shown in Figs. \ref{blackbox_m1} and \ref{blackbox_m2} where $\epsilon = 0$ and $\alpha = 0$ for the FGSM and PGD attack, respectively. In addition, the DAE significantly reduces the proportion of samples predicted with an infeasible power solution in each case over the considered perturbation magnitude range in comparison to no defense, adversarial training, and defensive distillation. For example, as shown in Fig. \ref{blackbox_m2}, a PGD adversarial attack at the maximum considered perturbation on samples with MR precoding results in an adversary success rate of approximately $70\%$. Adversarial training and defensive distillation result in an adversarial success rate to approximately $58\%$ and $82\%$, respectively, while our proposed DAE reduces the adversarial success rate to approximately $11\%$, which is approaching the expected performance of the DNN when no attacks are present (approximately $10\%$ of samples predicted with infeasible solutions). Lastly, we see that our proposed defense is able to mitigate FGSM attacks despite only being trained on PGD perturbations on one perturbation magnitude. This reinforces our proposed method's ability to provide robustness against other first order $l_{\infty}$-bounded adversarial attacks that the DAE was not trained to defend.

\subsection{Computational Efficiency} \label{sec:miti_eff}

Lastly, we discuss the computational feasibility of our proposed defense during both offline and online usage. Offline computational overhead is primarily associated with the training costs of the defense. During training, our proposed DAE takes, on average, $2.01$ secs. per epoch to train whereas adversarial training and defensive distillation require, on average, $3.10$ secs. per epoch. In all cases, these average training times assume that the defense only requires one perturbation bound for training. Advantageously, our proposed DAE defense can significantly mitigate a wide range of perturbation magnitudes when only trained on one bound (as described in Algorithm \ref{dae_const} and shown in Sec VI-B and VI-C), while adversarial training may require additional bounds leading to even higher offline computational costs. In addition, the computational cost of adversarial training and defensive distillation is dependent on the DNN architecture and its total number of trainable parameters (since they both require retraining) whereas the training cost of the DAE is independent of the DNN architecture. For example, the DAE trained to defend both Model 1 and Model 2 incur identical computational costs despite Model 2 having more parameters than Model 1. As a result, our proposed DAE can be implemented on high-parameter DNNs for strong adversarial attack mitigation without incurring high offline computational costs. 

Online computational costs correspond to the total time required at the BS to predict the power allocation to each UE in the cell. The forward propagation time through the DNN at the BS for power prediction on one sample (i.e., one particular distribution of UE positions) is $1.25 \times 10^{-3}$ secs. Forward propagation through the DAE prior to inputting the sample into the DNN adds an additional $0.87 \times 10^{-3}$ secs. for both Model 1 and Model 2, thus resulting in an online prediction time of our proposed defense of $2.12 \times 10^{-3}$ secs. Thus, with our proposed defense in place, the prediction time during online implementation remains very low and allows for rapid recalculation of power allocation as UEs quickly move within cells.






\section{Conclusion} \label{conclusion_sec}

Power allocation from base stations (BSs) to user equipments (UEs) is a vital task in maMIMO networks for low-interference and low-cost communications. Recent work has shown that deep neural networks (DNNs) can perform effective power allocation with reduced computational overhead in comparison to convex optimization techniques. Yet, DNNs are susceptible to adversarial attacks, which cause trained DNNs to predict infeasible power allocation solutions to UEs, thus depleting their effectiveness outweighing their computational efficiency. In this work, we proposed defending power allocation DNN models from adversarial attacks using denoising autoencoders (DAEs), where input sample were first forward propagated through the DAE to filter any potentially added adversarial attack before being inputted to the DNN for power allocation prediction. Our proposed method significantly improves and outperforms previously considered baselines for power allocation prediction in the presence of adversarial attacks, and it can be adopted in existing systems with relatively little computational overhead. Future work will consider the scalability of our work in very large federated machine learning networks \cite{ml_networks}, where layers of devices are used between BSs to train and defend power allocation DNNs.

\bibliography{references}

\begin{thebibliography}{10}
\providecommand{\url}[1]{#1}
\csname url@samestyle\endcsname
\providecommand{\newblock}{\relax}
\providecommand{\bibinfo}[2]{#2}
\providecommand{\BIBentrySTDinterwordspacing}{\spaceskip=0pt\relax}
\providecommand{\BIBentryALTinterwordstretchfactor}{4}
\providecommand{\BIBentryALTinterwordspacing}{\spaceskip=\fontdimen2\font plus
\BIBentryALTinterwordstretchfactor\fontdimen3\font minus
  \fontdimen4\font\relax}
\providecommand{\BIBforeignlanguage}[2]{{%
\expandafter\ifx\csname l@#1\endcsname\relax
\typeout{** WARNING: IEEEtran.bst: No hyphenation pattern has been}%
\typeout{** loaded for the language `#1'. Using the pattern for}%
\typeout{** the default language instead.}%
\else
\language=\csname l@#1\endcsname
\fi
#2}}
\providecommand{\BIBdecl}{\relax}
\BIBdecl

\bibitem{mimo_survey}
K.~Zheng, L.~Zhao, J.~Mei, B.~Shao, W.~Xiang, and L.~Hanzo, ``Survey of
  large-scale mimo systems,'' \emph{IEEE Communications Surveys \& Tutorials},
  vol.~17, no.~3, pp. 1738--1760, 2015.

\bibitem{mamimo1}
S.~Parkvall, E.~Dahlman, A.~Furuskar, and M.~Frenne, ``Nr: The new 5g radio
  access technology,'' \emph{IEEE Communications Standards Magazine}, vol.~1,
  no.~4, pp. 24--30, 2017.

\bibitem{mimo}
T.~L. Marzetta, ``Noncooperative cellular wireless with unlimited numbers of
  base station antennas,'' \emph{IEEE Transactions on Wireless Communications},
  vol.~9, no.~11, pp. 3590--3600, 2010.

\bibitem{trad_pa_1}
L.~Zhao, H.~Zhao, F.~Hu, K.~Zheng, and J.~Zhang, ``Energy efficient power
  allocation algorithm for downlink massive mimo with mrt precoding,'' in
  \emph{IEEE 78th VTC Fall}, 2013, pp. 1--5.

\bibitem{trad_pa_2}
Q.~Zhang, S.~Jin, M.~McKay, D.~Morales-Jimenez, and H.~Zhu, ``Power allocation
  schemes for multicell massive mimo systems,'' \emph{IEEE Transactions on
  Wireless Communications}, vol.~14, no.~11, pp. 5941--5955, 2015.

\bibitem{trad_pa_3}
T.~Van~Chien, E.~Björnson, and E.~G. Larsson, ``Joint power allocation and
  user association optimization for massive mimo systems,'' \emph{IEEE
  Transactions on Wireless Communications}, vol.~15, no.~9, pp. 6384--6399,
  2016.

\bibitem{trad_pa_4}
P.~Liu, S.~Jin, T.~Jiang, Q.~Zhang, and M.~Matthaiou, ``Pilot power allocation
  through user grouping in multi-cell massive mimo systems,'' \emph{IEEE
  Transactions on Communications}, vol.~65, no.~4, pp. 1561--1574, 2017.

\bibitem{trad_pa_5}
Y.~Dai and X.~Dong, ``Power allocation for multi-pair massive mimo two-way af
  relaying with linear processing,'' \emph{IEEE Transactions on Wireless
  Communications}, vol.~15, no.~9, pp. 5932--5946, 2016.

\bibitem{trad_pa_6}
C.~Sun, X.~Gao, and Z.~Ding, ``Bdma in multicell massive mimo communications:
  Power allocation algorithms,'' \emph{IEEE Transactions on Signal Processing},
  vol.~65, no.~11, pp. 2962--2974, 2017.

\bibitem{beyond_5g}
J.~Zhang, E.~Björnson, M.~Matthaiou, D.~W.~K. Ng, H.~Yang, and D.~J. Love,
  ``Prospective multiple antenna technologies for beyond 5g,'' \emph{IEEE
  Journal on Selected Areas in Communications}, vol.~38, no.~8, pp. 1637--1660,
  2020.

\bibitem{dl_in_mimo}
L.~Sanguinetti, A.~Zappone, and M.~Debbah, ``Deep learning power allocation in
  massive mimo,'' in \emph{52nd Asilomar Conference on Signals, Systems, and
  Computers}, 2018, pp. 1257--1261.

\bibitem{dl_in_mimo2}
Y.~Zhao, I.~G. Niemegeers, and S.~H. De~Groot, ``Power allocation in cell-free
  massive mimo: A deep learning method,'' \emph{IEEE Access}, vol.~8, pp.
  87\,185--87\,200, 2020.

\bibitem{dl_in_mimo3}
A.~Koc, M.~Wang, and T.~Le-Ngoc, ``Deep learning based multi-user power
  allocation and hybrid precoding in massive mimo systems,'' \emph{arXiv
  preprint arXiv:2201.12659}, 2022.

\bibitem{dl_in_mimo4}
L.~Luo, J.~Zhang, S.~Chen, X.~Zhang, B.~Ai, and D.~W.~K. Ng, ``Downlink power
  control for cell-free massive mimo with deep reinforcement learning,''
  \emph{IEEE Transactions on Vehicular Technology}, vol.~71, no.~6, pp.
  6772--6777, 2022.

\bibitem{adv_mimo1}
B.~R. Manoj, M.~Sadeghi, and E.~G. Larsson, ``Adversarial attacks on deep
  learning based power allocation in a massive mimo network,'' in \emph{IEEE
  ICC}, 2021, pp. 1--6.

\bibitem{adv_mimo2}
Q.~Liu, J.~Guo, C.-K. Wen, and S.~Jin, ``Adversarial attack on dl-based massive
  mimo csi feedback,'' \emph{Journal of Communications and Networks}, vol.~22,
  no.~3, pp. 230--235, 2020.

\bibitem{adv_mimo3}
P.~M. Santos, B.~R. Manoj, M.~Sadeghi, and E.~G. Larsson, ``Universal
  adversarial attacks on neural networks for power allocation in a massive mimo
  system,'' \emph{IEEE Wireless Communications Letters}, pp. 1--1, 2021.

\bibitem{adv_mimo4}
B.~Kim, Y.~Shi, Y.~E. Sagduyu, T.~Erpek, and S.~Ulukus, ``Adversarial attacks
  against deep learning based power control in wireless communications,'' in
  \emph{2021 IEEE Globecom Workshops}, 2021, pp. 1--6.

\bibitem{mimo_spoof}
L.~Xiao, T.~Chen, G.~Han, W.~Zhuang, and L.~Sun, ``Game theoretic study on
  channel-based authentication in mimo systems,'' \emph{IEEE Transactions on
  Vehicular Technology}, vol.~66, no.~8, pp. 7474--7484, 2017.

\bibitem{mimo_atk2}
B.~Akgun, M.~Krunz, and O.~Ozan~Koyluoglu, ``Vulnerabilities of massive mimo
  systems to pilot contamination attacks,'' \emph{IEEE Transactions on
  Information Forensics and Security}, vol.~14, no.~5, pp. 1251--1263, 2019.

\bibitem{mimo_atk3}
T.~Hou, S.~Bi, T.~Wang, Z.~Lu, Y.~Liu, S.~Misra, and Y.~Sagduyu, ``Muster:
  Subverting user selection in mu-mimo networks,'' in \emph{IEEE INFOCOM 2022 -
  IEEE Conference on Computer Communications}, 2022, pp. 140--149.

\bibitem{amc_dl}
T.~J. O’Shea, T.~Roy, and T.~C. Clancy, ``Over-the-air deep learning based
  radio signal classification,'' \emph{IEEE Journal of Selected Topics in
  Signal Processing}, vol.~12, no.~1, pp. 168--179, 2018.

\bibitem{sl1}
M.~Arnold, J.~Hoydis, and S.~t. Brink, ``Novel massive mimo channel sounding
  data applied to deep learning-based indoor positioning,'' in \emph{12th
  International ITG Conference on Systems, Communications and Coding}, 2019,
  pp. 1--6.

\bibitem{sl2}
S.~D. Bast, A.~P. Guevara, and S.~Pollin, ``Csi-based positioning in massive
  mimo systems using convolutional neural networks,'' in \emph{2020 IEEE
  VTC2020-Spring}, 2020, pp. 1--5.

\bibitem{chan_est}
F.~Liang, C.~Shen, and F.~Wu, ``An iterative bp-cnn architecture for channel
  decoding,'' \emph{IEEE Journal of Selected Topics in Signal Processing},
  vol.~12, no.~1, pp. 144--159, 2018.

\bibitem{adv_in_wc}
D.~Adesina, C.-C. Hsieh, Y.~E. Sagduyu, and L.~Qian, ``Adversarial machine
  learning in wireless communications using rf data: A review,'' \emph{arXiv
  preprint arXiv:2012.14392}, 2020.

\bibitem{iot_adv_atk}
Z.~Bao, Y.~Lin, S.~Zhang, Z.~Li, and S.~Mao, ``Threat of adversarial attacks on
  dl-based iot device identification,'' \emph{IEEE Internet of Things Journal},
  vol.~9, no.~11, pp. 9012--9024, 2022.

\bibitem{amc_adv_atk1}
M.~Sadeghi and E.~G. Larsson, ``Adversarial attacks on deep-learning based
  radio signal classification,'' \emph{IEEE Wireless Communications Letters},
  vol.~8, no.~1, pp. 213--216, 2018.

\bibitem{amc_adv_atk2}
Y.~{Lin}, H.~{Zhao}, Y.~{Tu}, S.~{Mao}, and Z.~{Dou}, ``Threats of adversarial
  attacks in dnn-based modulation recognition,'' in \emph{IEEE INFOCOM}, 2020,
  pp. 2469--2478.

\bibitem{amc_adv_atk3}
B.~{Kim}, Y.~E. {Sagduyu}, K.~{Davaslioglu}, T.~{Erpek}, and S.~{Ulukus},
  ``Over-the-air adversarial attacks on deep learning based modulation
  classifier over wireless channels,'' in \emph{54th Annual CISS}, 2020, pp.
  1--6.

\bibitem{amc_adv_atk4}
A.~Berian, K.~Staab, N.~Teku, G.~Ditzler, T.~Bose, and R.~Tandon, ``Adversarial
  filters for secure modulation classification,'' \emph{arXiv preprint
  arXiv:2008.06785}, 2020.

\bibitem{amc_adv_atk5}
Y.~Lin, H.~Zhao, X.~Ma, Y.~Tu, and M.~Wang, ``Adversarial attacks in modulation
  recognition with convolutional neural networks,'' \emph{IEEE Transactions on
  Reliability}, vol.~70, no.~1, pp. 389--401, 2021.

\bibitem{adversary_atk}
Y.~E. Sagduyu, Y.~Shi, and T.~Erpek, ``Iot network security from the
  perspective of adversarial deep learning,'' in \emph{16th Annual IEEE
  International Conference on Sensing, Communication, and Networking (SECON)},
  2019, pp. 1--9.

\bibitem{eve1}
Y.~Shi, Y.~E. Sagduyu, T.~Erpek, K.~Davaslioglu, Z.~Lu, and J.~H. Li,
  ``Adversarial deep learning for cognitive radio security: Jamming attack and
  defense strategies,'' in \emph{IEEE ICC Workshops}, 2018, pp. 1--6.

\bibitem{eve2}
T.~Erpek, Y.~E. Sagduyu, and Y.~Shi, ``Deep learning for launching and
  mitigating wireless jamming attacks,'' \emph{IEEE Transactions on Cognitive
  Communications and Networking}, vol.~5, no.~1, pp. 2--14, 2019.

\bibitem{eve3}
M.~Z. Hameed, A.~György, and D.~Gündüz, ``The best defense is a good
  offense: Adversarial attacks to avoid modulation detection,'' \emph{IEEE
  Transactions on Information Forensics and Security}, vol.~16, pp. 1074--1087,
  2021.

\bibitem{ae_pt}
S.~{Kokalj-Filipovic}, R.~{Miller}, N.~{Chang}, and C.~L. {Lau}, ``Mitigation
  of adversarial examples in rf deep classifiers utilizing autoencoder
  pre-training,'' in \emph{ICMCIS}, 2019, pp. 1--6.

\bibitem{adv_amc_detection}
L.~Zhang, S.~Lambotharan, G.~Zheng, B.~AsSadhan, and F.~Roli, ``Countermeasures
  against adversarial examples in radio signal classification,'' \emph{IEEE
  Wireless Communications Letters}, vol.~10, no.~8, pp. 1830--1834, 2021.

\bibitem{ade}
R.~Sahay, C.~G. Brinton, and D.~J. Love, ``A deep ensemble-based wireless
  receiver architecture for mitigating adversarial attacks in automatic
  modulation classification,'' \emph{IEEE Transactions on Cognitive
  Communications and Networking}, vol.~8, no.~1, pp. 71--85, 2022.

\bibitem{cv_review}
N.~Akhtar and A.~Mian, ``Threat of adversarial attacks on deep learning in
  computer vision: A survey,'' \emph{IEEE Access}, vol.~6, pp.
  14\,410--14\,430, 2018.

\bibitem{dae}
R.~Sahay, R.~Mahfuz, and A.~E. Gamal, ``Combatting adversarial attacks through
  denoising and dimensionality reduction: A cascaded autoencoder approach,'' in
  \emph{53rd Annual CISS}, 2019, pp. 1--6.

\bibitem{dunet}
F.~Liao, M.~Liang, Y.~Dong, T.~Pang, X.~Hu, and J.~Zhu, ``Defense against
  adversarial attacks using high-level representation guided denoiser,'' in
  \emph{Proc. of the IEEE CVPR}, June 2018.

\bibitem{pgd_adv_trn}
A.~Madry, A.~Makelov, L.~Schmidt, D.~Tsipras, and A.~Vladu, ``Towards deep
  learning models resistant to adversarial attacks,'' in \emph{International
  Conference on Learning Representations}, 2018.

\bibitem{fgsm}
I.~J. Goodfellow, J.~Shlens, and C.~Szegedy, ``Explaining and harnessing
  adversarial examples,'' \emph{arXiv:1412.6572}, 2014.

\bibitem{adv_trn_sahay}
R.~Sahay, D.~J. Love, and C.~G. Brinton, ``Robust automatic modulation
  classification in the presence of adversarial attacks,'' in \emph{2021 55th
  Annual CISS}, 2021, pp. 1--6.

\bibitem{mimo_adv_atk_def}
L.~Zhang, S.~Lambotharan, and G.~Zheng, ``A countermeasure against adversarial
  attacks on power allocation in a massive mimo network,'' in \emph{2022 IEEE
  Symposium on Wireless Technology \& Applications (ISWTA)}, 2022, pp. 13--16.

\bibitem{mimo_adv_trn}
B.~R. Manoj, M.~Sadeghi, and E.~G. Larsson, ``Downlink power allocation in
  massive mimo via deep learning: Adversarial attacks and training,''
  \emph{IEEE Transactions on Cognitive Communications and Networking}, vol.~8,
  no.~2, pp. 707--719, 2022.

\bibitem{adv_trn_ovft}
L.~Rice, E.~Wong, and J.~Z. Kolter, ``Overfitting in adversarially robust deep
  learning,'' in \emph{ICML}, 2020.

\bibitem{obf_grd}
A.~Athalye, N.~Carlini, and D.~A. Wagner, ``Obfuscated gradients give a false
  sense of security: Circumventing defenses to adversarial examples,'' in
  \emph{ICML}, 2018, pp. 274--283.

\bibitem{bb_justification}
Y.~{Sagduyu}, Y.~{Shi}, and T.~{Erpek}, ``Adversarial deep learning for
  over-the-air spectrum poisoning attacks,'' \emph{IEEE Transactions on Mobile
  Computing}, 2019.

\bibitem{mamimo_book}
\BIBentryALTinterwordspacing
E.~Björnson, J.~Hoydis, and L.~Sanguinetti, ``Massive mimo networks: Spectral,
  energy, and hardware efficiency,'' \emph{Foundations and Trends® in Signal
  Processing}, vol.~11, no. 3-4, pp. 154--655, 2017. [Online]. Available:
  \url{http://dx.doi.org/10.1561/2000000093}
\BIBentrySTDinterwordspacing

\bibitem{mr_precoding}
T.~L. Marzetta, \emph{Fundamentals of massive MIMO}.\hskip 1em plus 0.5em minus
  0.4em\relax Cambridge University Press, 2016.

\bibitem{mamimo2}
E.~Björnson, J.~Hoydis, and L.~Sanguinetti, ``Massive mimo has unlimited
  capacity,'' \emph{IEEE Transactions on Wireless Communications}, vol.~17,
  no.~1, pp. 574--590, 2018.

\bibitem{spoofing}
M.~L. Psiaki and T.~E. Humphreys, ``Gnss spoofing and detection,''
  \emph{Proceedings of the IEEE}, vol. 104, no.~6, pp. 1258--1270, 2016.

\bibitem{grad1}
X.~Yuan, P.~He, Q.~Zhu, and X.~Li, ``Adversarial examples: Attacks and defenses
  for deep learning,'' \emph{IEEE Transactions on Neural Networks and Learning
  Systems}, vol.~30, no.~9, pp. 2805--2824, 2019.

\bibitem{grad2}
Y.~Dong, F.~Liao, T.~Pang, H.~Su, J.~Zhu, X.~Hu, and J.~Li, ``Boosting
  adversarial attacks with momentum,'' \emph{IEEE/CVF Conference on Computer
  Vision and Pattern Recognition}, pp. 9185--9193, 2018.

\bibitem{pgd}
A.~Madry, A.~Makelov, L.~Schmidt, D.~Tsipras, and A.~Vladu, ``Towards deep
  learning models resistant to adversarial attacks,'' \emph{International
  Conference on Learning Representations (ICLR)}, 2018.

\bibitem{transfer1}
Y.~Liu, X.~Chen, C.~Liu, and D.~Song, ``Delving into transferable adversarial
  examples and black-box attacks,'' in \emph{International Conference on
  Learning Representations (ICLR)}, 2017.

\bibitem{transfer2}
F.~Tramer, N.~Papernot, I.~Goodfellow, D.~Boneh, and P.~McDaniel, ``The space
  of transferable adversarial examples,'' \emph{arXiv preprint
  arXiv:1704.03453}, 2017.

\bibitem{transfer3}
N.~Papernot, P.~McDaniel, and I.~Goodfellow, ``Transferability in machine
  learning: from phenomena to black-box attacks using adversarial samples,''
  \emph{arXiv preprint arXiv:1605.07277}, 2016.

\bibitem{resnet}
K.~He, X.~Zhang, S.~Ren, and J.~Sun, ``Deep residual learning for image
  recognition,'' \emph{IEEE Conference on Computer Vision and Pattern
  Recognition}, pp. 770--778, 2016.

\bibitem{dl_book}
I.~Goodfellow, Y.~Bengio, and A.~Courville, \emph{Deep Learning}.\hskip 1em
  plus 0.5em minus 0.4em\relax MIT Press, 2016,
  \url{http://www.deeplearningbook.org}.

\bibitem{def_dist}
N.~Papernot, P.~McDaniel, X.~Wu, S.~Jha, and A.~Swami, ``Distillation as a
  defense to adversarial perturbations against deep neural networks,'' in
  \emph{2016 IEEE Symposium on Security and Privacy (SP)}, 2016, pp. 582--597.

\bibitem{def_dist_not_rob}
N.~Carlini and D.~Wagner, ``Defensive distillation is not robust to adversarial
  examples,'' \emph{arXiv preprint arXiv:1607.04311}, 2016.

\bibitem{ml_networks}
S.~Hosseinalipour, S.~S. Azam, C.~G. Brinton, N.~Michelusi, V.~Aggarwal, D.~J.
  Love, and H.~Dai, ``Multi-stage hybrid federated learning over large-scale
  d2d-enabled fog networks,'' \emph{IEEE Transactions on Networking}, pp.
  1--16, 2022.

\end{thebibliography}

\bibliographystyle{IEEEtran}

\end{document}